\documentclass{nature_mod}

\usepackage{url}
\usepackage{epsfig}
\usepackage{graphicx}
\usepackage{xcolor}
\usepackage{caption}
\usepackage{hyperref}
\usepackage{lineno}
\usepackage{aas_macros}
\usepackage{threeparttable}
\usepackage{multibib}
\newcites{methods}{Methods References}
\usepackage{amssymb,amsmath,caption}

\newcommand{\pasa}{Publications of the Astronomical Society of Australia}
\newcommand{\raa}{Research in Astronomy and Astrophysics}

\long\def\symbolfootnote[#1]#2{\begingroup%
\def\thefootnote{\fnsymbol{footnote}}\footnote[#1]{#2}\endgroup} 
\title{An invariant energy release hierarchy in a repeating fast radio burst}
\author{X. Yang$^{1}$\thanks{These authors contributed equally to this work.}, S. B. Zhang$^{1,2 \ast}$, Y. Li$^{1,3,4 \ast}$, D. Xiao$^{1,3,4}$\thanks{Email: dxiao@pmo.ac.cn}, W. L. Zhang$^{1,3}$, J.-J. Wei$^{1,3}$, J.-J. Geng$^{1,3}$, J.-S. Wang$^{5}$, Y. P. Yang$^{6}$, F. Y. Wang$^{7,8}$, X. F. Wu$^{1,3}$\thanks{Email: xfwu@pmo.ac.cn}, Z. G. Dai$^{9}$\thanks{Email: daizg@ustc.edu.cn}}

\begin{document}

\maketitle

\begin{affiliations}
 \item Purple Mountain Observatory, Chinese Academy of Sciences, Nanjing 210023, China
 \item CSIRO Space and Astronomy, Australia Telescope National Facility, P.O. Box 76, Epping, NSW 1710, Australia
 \item School of Astronomy and Space Sciences, University of Science and Technology of China, Hefei 230026, China
 \item State Key Laboratory of Radio Astronomy and Technology, Purple Mountain Observatory, Chinese Academy of Sciences, 10 Yuanhua Road, Nanjing 210023, China
 \item Tsung-Dao Lee Institute, Shanghai Jiao Tong University, Shanghai 200240, China
 \item South-Western Institute for Astronomy Research, Yunnan Key Laboratory of Survey Science, Yunnan University, Kunming 650091, China
 \item School of Astronomy and Space Science, Nanjing University, Nanjing 210093, People’s Republic of China
 \item Key Laboratory of Modern Astronomy and Astrophysics, Nanjing University, Nanjing 210093, People’s Republic of China
 \item Department of Astronomy, University of Science and Technology of China, Hefei 230026, China

\end{affiliations}

\section*{Abstract}
Fast radio bursts (FRBs) are luminous millisecond radio transients whose physical origin remains unsettled. A key diagnostic is whether their burst-energy distributions retain characteristic physical scales that are intrinsic and temporally stable within an individual engine.
Here we report a 3.2-year monitoring campaign of the hyperactive repeater FRB~20220529 with FAST and Parkes, yielding more than 1,300 bursts spanning nearly five orders of magnitude in spectral energy density. The cumulative burst-rate distribution is described by an exponential-plus-power-law (EXP+PL) form, linking a low-energy exponential component with characteristic scale \(E_0\) to a scale-free bright-end tail. This scale remains invariant despite the burst rate declining by more than an order of magnitude, revealing a stable dissipation scale decoupled from the source's macroscopic trigger activity.
Within a magnetar interpretation, this phenomenology is consistent with localized sub-critical reconnection episodes coexisting with plasmoid-mediated magnetic avalanches in a twisted magnetosphere. The invariant \(E_0\) constrains the dissipation region to the inner-to-middle magnetosphere and reveals a robust energy-release hierarchy beneath the variable activity of repeating FRBs, providing an observational benchmark for relativistic reconnection in an ultra-magnetized neutron-star environment.

\bigskip

Whether burst-like energy-release systems are governed by scale-free cascades alone or by processes that retain characteristic physical scales is a question that cuts across disciplines, from solar and stellar flares to earthquakes. Repeating fast radio bursts (FRBs)\cite{Lorimer,frbreview} provide a compact-object laboratory to address this problem. Previous FRB studies have revealed broad scale-free behavior and, in some cases, candidate characteristic energies through power-law, broken-power-law or multi-component descriptions \cite{Wei_2021ApJ,ZhangGQ2021ApJ,Kirsten_2024NatAs,Wu_2025ApJ,li121102,zhang2024bimodal,Konijn2024,songbo2025}. The key unresolved question is whether these scales are persistent properties of an individual engine, rather than features produced by sampling, activity state or observational selection. A stringent test is temporal stability: an intrinsic characteristic scale should persist as the source activity evolves, whereas a fitting artifact or transient state-dependent feature need not.

To test this hypothesis, we carried out a 3.2-year monitoring campaign of the hyperactive repeater FRB~20220529 with the Five-hundred-meter Aperture Spherical radio Telescope (FAST) and the Parkes 64-m radio telescope. From more than 1,300 detected bursts spanning nearly five orders of magnitude in spectral energy density, we identify an exponential-plus-power-law (EXP+PL) energy distribution whose characteristic low-energy scale \(E_0\) remains invariant despite the burst rate declining by more than an order of magnitude. We interpret this two-component structure within a magnetar reconnection framework and use the stable characteristic scale to constrain the magnetospheric emission region.

\section*{A stable characteristic energy scale in a two-component distribution}
\begin{figure*}
    \begin{center}
    \begin{tabular}{cc}
    \includegraphics[width=75mm]{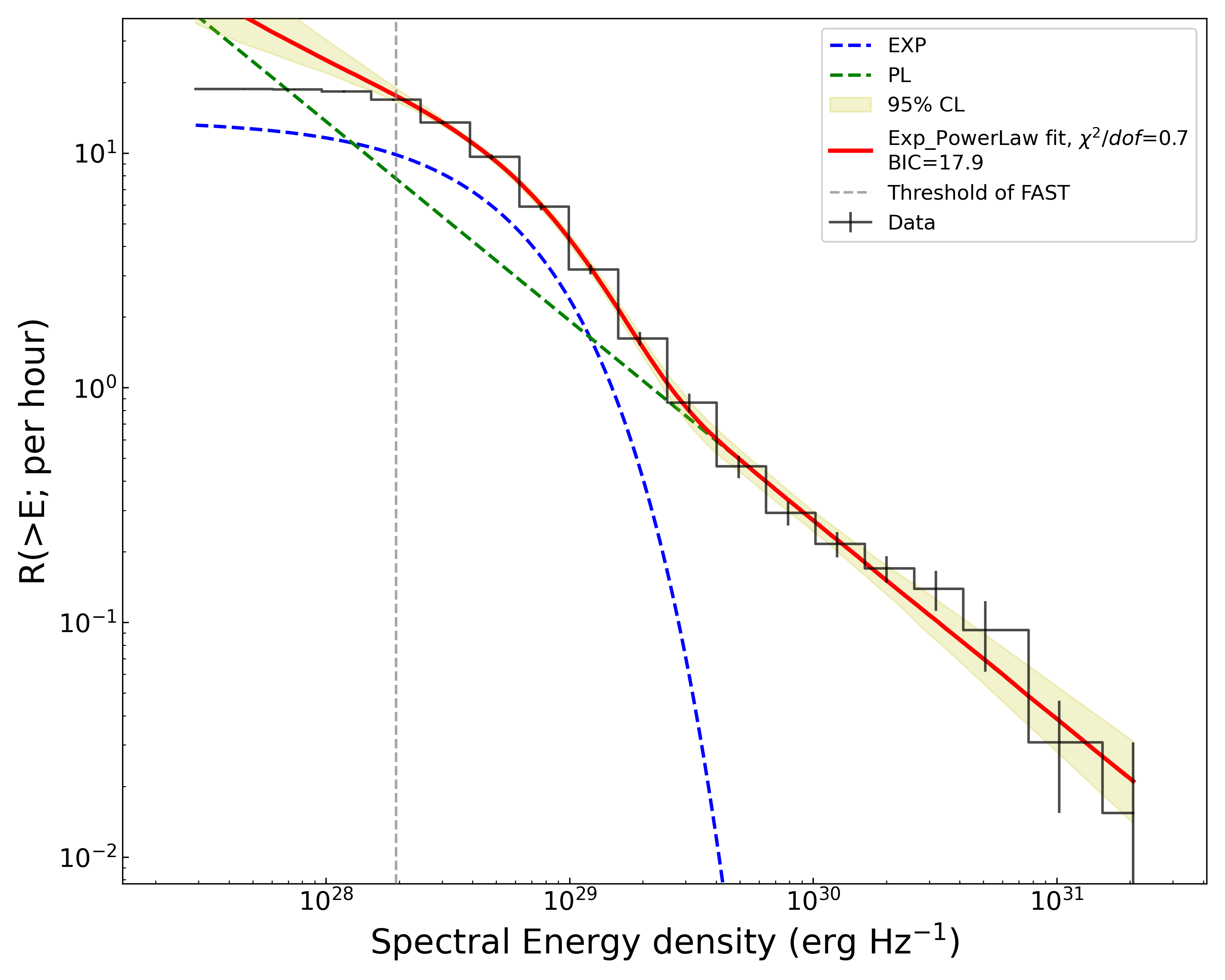} & 
    \includegraphics[width=75mm]{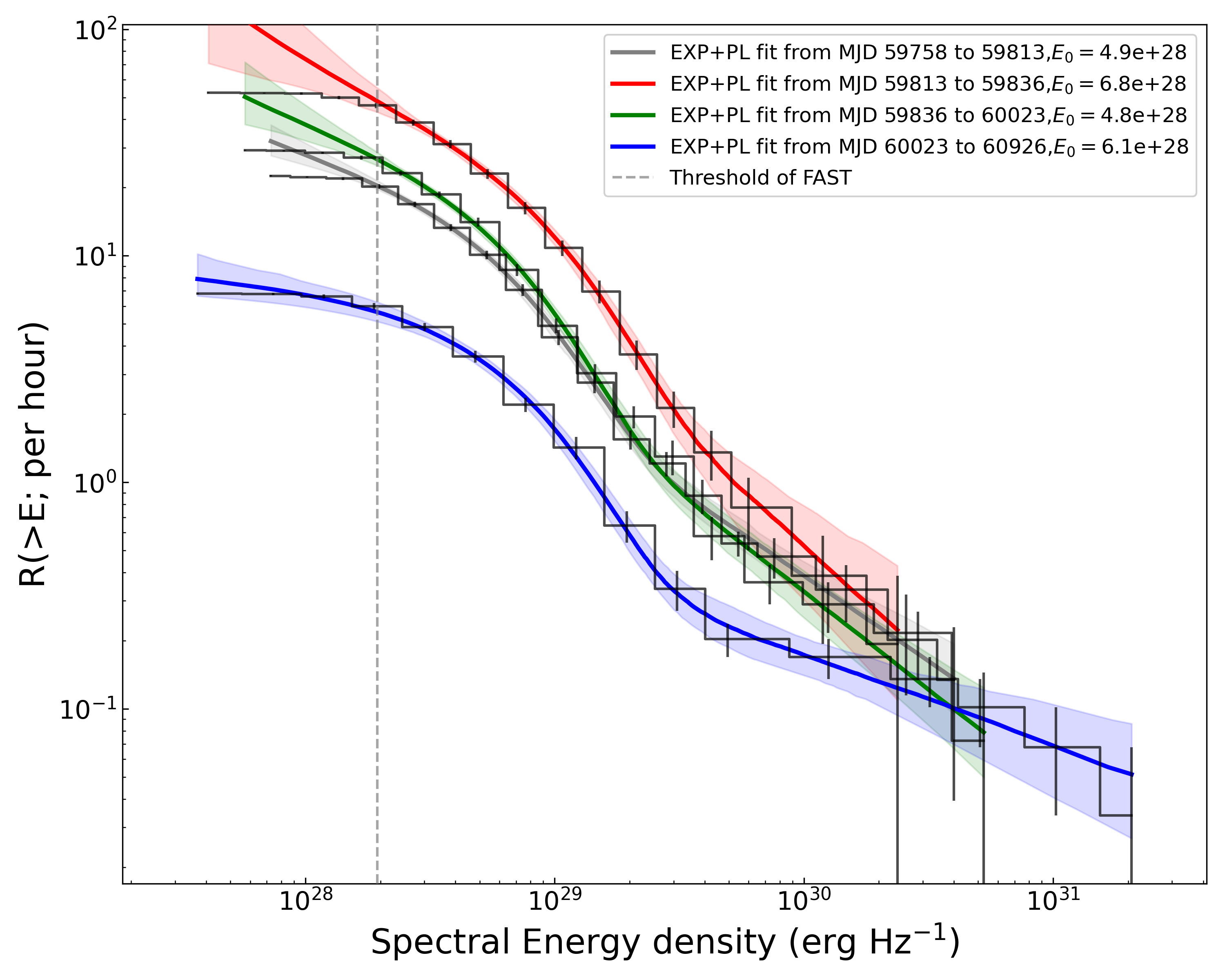}\\
    \end{tabular}
    \caption{The left panel shows the cumulative burst-rate distributions for FAST observations (full sample). The right panel shows the epoch-resolved stability of the EXP+PL energy distribution. Colored lines represent the best-fit exponential-plus-power-law (EXP+PL) models, with shaded regions indicating 95\% credible intervals. Vertical dashed lines mark the 95\% completeness threshold. Despite a substantial drop in the global burst rate, the characteristic exponential scale \(E_0\) remains stable across all epochs.}
    \label{figure:frb_spc}
    \end{center}
    \end{figure*}

We monitored FRB~20220529 from 22 June 2022 to 8 September 2025 using the FAST and Parkes telescopes (Fig.~\ref{figure:frb_obs}).
To quantitatively establish the energy distribution, we first analyzed the densely sampled FAST dataset, which covers a spectral energy density range of $2.3\times10^{27} - 2.6\times10^{31}$ erg~Hz$^{-1}$.
As shown in Table~\ref{table:energy_fit}, across the FAST observed energy range, statistical comparisons reveal that single-component models (purely power-law or purely exponential) are strongly disfavored (see Methods). Instead, the exponential-plus-power-law (EXP+PL) model provides the best description of the data, as evidenced by its significantly lower reduced chi-squared statistic ($\chi^2/\mathrm{dof}$) and Bayesian Information Criterion (BIC)\cite{Clauset2009}.
This model decomposes the burst population into a low-energy component characterized by an e-folding scale \(E_0\) and a scale-free bright-end tail (Fig.~\ref{figure:frb_spc}). 
%Similar exponential-to-power-law transitions are observed in solar and stellar flare energy distributions, where they are often interpreted as signatures of thresholded dissipation regimes~\cite{Aschwanden2021}.

The fitting was restricted to the energy range above the 95\% detection completeness threshold to avoid observational bias from incomplete sampling. To further verify that the high-energy power-law tail is not a statistical fluctuation of the FAST sample, we complemented the analysis with independent ultra-wideband observations from the Parkes telescope, covering the same 3.2-year period. The Parkes data were reserved exclusively for cross-validation of the high-energy component rather than contributing to the core fitting.
Extending the distribution to rarer, brighter events, the Parkes dataset is well-described by a single power-law model (left panel of Extended Data Fig.~\ref{fig:compare0912}). The consistency between the Parkes high-energy bursts and the FAST-only power-law fit reinforces the robustness of the high-energy component. This agreement extends the dynamic range of the energy distribution to nearly five orders of magnitude, confirming that the bright end continues as a scale-free component rather than terminating in an exponential cutoff. An independent Tsallis \(q\)-Gaussian analysis of the burst-energy return distribution further corroborates this picture: the full sample departs from scale invariance, as expected when a characteristic energy scale is present, whereas the high-energy subset (\(E_{\nu}>5\times10^{29}\)~erg~Hz\(^{-1}\)) approaches scale-invariant statistics consistent with an avalanche-like cascade (see Methods).

The central finding is not merely the identification of a two-component distribution, but the remarkable temporal invariance of its characteristic energy scale \(E_0\). To assess the temporal stability of this dissipation hierarchy, we divided the FAST sample into four chronological epochs. For each epoch, fitting was performed above the 95\% completeness threshold. The EXP+PL morphology persists throughout the 3.2-year baseline despite a strong decline in the global burst rate (Fig.~\ref{figure:frb_spc}). Most notably, \(E_0\) remains confined to \((4.83\text{--}6.82)\times10^{28}\) erg~Hz\(^{-1}\), even as the component normalizations vary with activity level and are substantially reduced in the late-time, low-activity epoch (Table~\ref{tab:4epoch_fit}). Thus, $E_0$ varies by less than a factor of 1.5 while the global burst rate drops by more than an order of magnitude, indicating that the characteristic dissipation scale is decoupled from the macroscopic trigger frequency.

The high-energy tail shows a similar stability during the burst-rich part of the monitoring campaign: the power-law index remains broadly consistent across the first three epochs and flattens only in the final, low-activity stage. As a robustness check, fixing the Epoch~4 power-law index to the full-sample value, $\gamma=0.85$, yields $E_0=6.07\times10^{28}$ erg Hz$^{-1}$, consistent with the other epochs. This indicates that the apparent late-time flattening is likely driven by the limited number of bright bursts in the low-activity regime and does not alter the inference of a stable exponential scale.

\section*{Comparison with other hyperactive repeaters}
To investigate whether this dual-mode energy release may recur in other repeating FRB engines, we examined the burst statistics of other hyperactive repeaters. Previous studies have identified deviations from a single power-law energy distribution in several active repeaters, including FRB~20121102, FRB~20200120E, and FRB~20220912A, typically characterized by broken-power-law or bimodal parameterizations \cite{li121102,zhang2024bimodal,Konijn2024,songbo2025}. The EXP+PL model, first proposed in this work, provides a physically distinct decomposition of the burst population, associating the exponential and power-law components with distinct dissipation regimes. Notably, the observed cumulative burst-rate distribution of FRB~20220912A \cite{Konijn2024} displays a phenomenologically similar exponential-to-power-law transition over a comparable broad energy range (Extended Data Fig.~\ref{fig:compare0912}; see Methods). Despite differences in observing setups, frequency coverage, and selection functions, the two sources show a closely similar qualitative morphology.

This phenomenological convergence argues against a purely source-specific or instrumental origin for the observed hierarchy. However, the broader repeating-FRB population has not yet been systematically surveyed for dual-mode energy distributions, and the available sample statistics remain limited outside a few well-monitored sources \cite{CHIME2021,Nimmo2023,CHIME2026}. While several other active repeaters currently lack sufficient burst statistics or dynamic range to resolve a dual-mode structure (see Methods), the alignment between FRB~20220529 and FRB~20220912A provides an independent indication that such an energy hierarchy may recur among hyperactive repeaters. Whether the characteristic scale $E_0$ is similarly invariant in FRB~20220912A remains an open question testable with epoch-resolved analysis.

\section*{A thresholded reconnection hierarchy in a magnetar magnetosphere}

The coexistence of a characteristic low-energy scale and a scale-free high-energy tail constrains models of the burst energetics. In particular, the temporal invariance of \(E_0\) over 3.2 years, despite a decline of more than an order of magnitude in the burst rate, indicates that the two-component structure is an intrinsic property of the engine rather than a transient observational feature. In a magnetar reconnection picture, this behaviour is naturally accommodated if the low-energy component is set by a relatively stable local dissipation scale, whereas the burst rate is controlled mainly by how often such reconnection sites are activated. The source can then fade by triggering fewer reconnection episodes without shifting the characteristic scale of each localized event.

Within this framework, magnetic energy stored in a twisted magnetosphere can be released through intermittent current-sheet formation and reconnection \cite{Thompson2002,Beloborodov2009,Parfrey2013}. The exponential component can be associated with localized reconnection episodes whose effective branching remains subcritical, so that the dissipation is confined to small participating volumes. By contrast, the power-law tail is consistent with independent realizations of near-critical plasmoid-mediated avalanches whose effective participating length or volume has no preferred scale \cite{Samtaney2009,Uzdensky2010,HuangBhattacharjee2010,HuangBhattacharjee2012,LoureiroUzdensky2016}. Successive bursts need not be causally connected to one another; rather, each burst can be viewed as a different realization drawn from the same underlying avalanche-size distribution. The relevant criticality is therefore associated with the effective branching and participating scale of an individual reconnection episode, rather than with the global Lundquist number of the magnetospheric current sheet alone (See Supplementary). This interpretation constrains the magnetic energy-release hierarchy, while leaving the detailed conversion into coherent radio emission model dependent \cite{Philippov2019,Lyubarsky2021}.

\begin{figure*}
		\centering
		\includegraphics[width=0.85\textwidth]{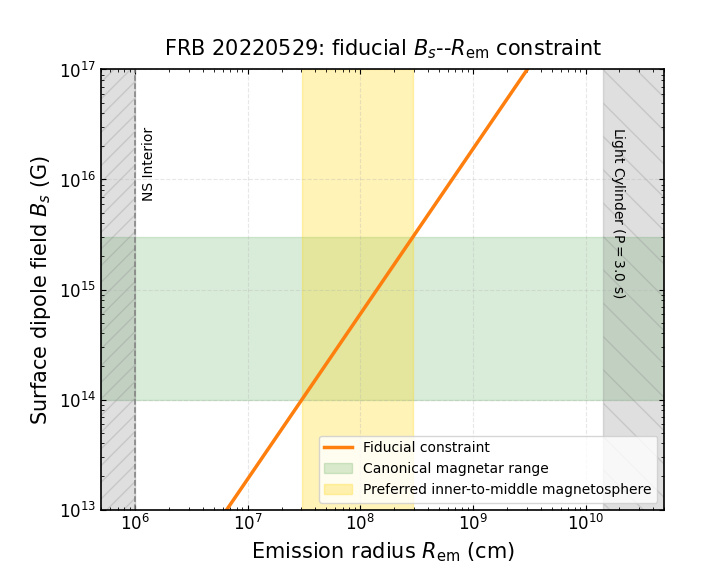}
		\caption{Physical constraints on the magnetar engine parameters. The derived locus in the surface dipole field (\(B_s\)) and emission radius (\(R_{\mathrm{em}}\)) plane is inferred from the stable full-sample exponential scale \(E_0\). For canonical magnetar field strengths (green band), the \(E_0\)-anchored constraint places the characteristic localized dissipation layer in the inner-to-middle magnetosphere (orange band). Hatched regions indicate the neutron-star interior and the region beyond the light cylinder for a 3-second spin period. The dependence on the composite nuisance parameter and the additional active-volume constraints are presented in Supplementary Note~3.}
		\label{fig:constraints_main}
	\end{figure*}

The stable exponential scale, \(E_0=5.68\times10^{28}\) erg Hz\(^{-1}\) for the full FAST sample, provides the quantitative anchor for the local magnetospheric energy reservoir. Interpreting this scale as the characteristic energy of localized subcritical reconnection episodes constrains the emission site to \(R_{\mathrm{em}}\sim10^8\) cm for canonical magnetar field strengths, placing the relevant dissipation layer in the inner-to-middle magnetosphere (Fig.~\ref{fig:constraints_main}). The required effective active volume is far larger than the minimal scale expected for a single critical plasmoid, indicating that the observed low-energy scale is associated with a mesoscopic reconnecting region rather than an isolated microscopic dissipation site (See Supplementary). The brightest bursts then correspond to the scale-free avalanche branch, in which a subset of events accesses larger participating reconnection scales.

Alternative scenarios involving spatially independent emission sites or external triggers are not excluded, but they must still explain why the same characteristic scale persists as the source activity declines, and how a scale-bearing low-energy component connects naturally to a scale-free bright-end tail without fine tuning.

\section*{Broader implications and testable predictions}
Establishing a stable dissipation hierarchy bridges burst statistics with relativistic magnetic reconnection in an extreme compact-object environment. It elevates burst-energy distributions from phenomenological descriptions to stability tests of the underlying engine. In this sense, repeating FRBs provide a new observational route for probing multi-scale plasma dynamics in ultra-magnetized neutron-star magnetospheres.

This framework yields several testable extensions beyond the present dataset. First, in other long-lived active repeaters, or in future monitoring of FRB~20220529, the characteristic exponential scale \(E_0\) should remain more stable than the rate normalizations over long timescales, even as the global burst rate varies. Second, the observed characteristic scale and the relative visibility of the exponential and power-law branches may depend on observing frequency. If different radio bands sample different magnetospheric radii or radiative-efficiency regimes, broadband observations should reveal systematic changes in \(E_0\) or in the apparent balance between the two components, providing a probe of the emission geometry. Third, phases of enhanced burst activity may be accompanied by changes in the local magneto-ionic environment, including variations in rotation measure, dispersion measure (DM) or polarization properties. Such correlations would test whether the same magnetic restructuring that triggers bursts also perturbs the surrounding plasma.

A key future test will be uniform, epoch-resolved EXP+PL modeling of a sample of hyperactive repeaters over multi-year baselines, with high cadence, broad frequency coverage and well-characterized completeness. Such observations will determine whether stable thresholded energy hierarchies are common among active repeating FRBs, or instead restricted to a subset of long-lived hyperactive engines.

\clearpage

\clearpage
%% MAIN PAPER REFERENCES

\clearpage

	\begin{table*}
		\caption{\bf Burst energy ranges and fitting results of energy distribution for FRB 20220529. The fitting results are obtained from FAST observation sample.}
		\renewcommand\arraystretch{1.1}
		\begin{center}
			\begin{threeparttable}
				\begin{tabular}{lcc}
					\hline
					{\bf Property} & \multicolumn{2}{c}{\bf Measurement} \\
					\hline
					% Energy ranges (单独占一行跨列显示)
					FAST energy density~(erg~Hz$^{-1}$) & \multicolumn{2}{c}{$2.3\times10^{27} - 2.6\times10^{31}$} \\
					Parkes energy density~(erg~Hz$^{-1}$) & \multicolumn{2}{c}{$6.5\times10^{29} - 8.6\times10^{31}$} \\
					\hline
					{\bf Model} & $\mathbf{\chi^{2}/dof}$ & {\bf BIC} \\
					\hline
					
					Single power law & 7.3 & 92.3 \\
					Single exponential & 34.3 & 416.4 \\
					[0.5em]
					Broken power law & 6.8 & 78.29 \\
                    Smoothly broken power law & 4.3 & 52.3 \\
					Double exponential & 4.9 & 59.4 \\
					Exponential + power law & 0.7 & 17.9 \\
					\hline
				\end{tabular}
				%   \begin{tablenotes}
					%        \footnotesize
					%        \item[a]Corrected to .
					
					%     \end{tablenotes}
			\end{threeparttable}
		\end{center}
		\label{table:energy_fit}
	\end{table*}
	
	\begin{table*}
		\centering
		\caption{Best-fit parameters of the cumulative burst-rate distribution for four epochs and the full burst sample of FRB 20220529 from FAST observation. 
			The distribution is modeled as $R(>E) = A \exp\left(-E/E_0\right) + B \left(E/10^{30}\ \mathrm{erg\ Hz^{-1}}\right)^{-\gamma}$, where $R(>E)$ is measured in bursts~h$^{-1}$ and $A$ and $B$ are rate normalizations.
			Each epoch contains approximately 300 bursts. Here \(E_{\nu,c}\) is the fitted rate-crossover scale defined in Methods.}
		\label{tab:4epoch_fit}
        \scalebox{1.0}{ 
		\begin{tabular}{lccccccc}
			\hline
			Epoch         & MJD Start & MJD End & $A$    & $E_0$ ($\mathrm{erg\ Hz^{-1}}$) & $B$   & $\gamma$  & $E_{\nu,c}$ ($\rm erg\ Hz^{-1}$)\\
			\hline
			Full sample   & 59758       & 60925     & 13.79  & $5.68\times10^{28}$             & 0.27  & 0.85 & $1.27\times10^{28}$\\
			Epoch 1       & 59758     & 59812   & 19.74   & $4.95\times10^{28}$             & 0.39  & 0.74 & $5.84\times10^{27}$\\
			Epoch 2       & 59812     & 59835   & 30.92 & $6.82\times10^{28}$             & 0.51  & 0.99 & $2.19\times10^{28}$\\
			Epoch 3       & 59835     & 60022   & 24.89  & $4.83\times10^{28}$             & 0.33  & 0.87 & $8.50\times10^{27}$\\
			Epoch 4       & 60022     & 60925   & 6.63    & $6.10\times10^{28}$             & 0.17  & 0.39 & $1.83\times10^{29}$\\

			\hline
		\end{tabular}
        }
	\end{table*}
	
	%% FIGURES

	\begin{figure*}
		\centering
		\includegraphics[width=160mm]{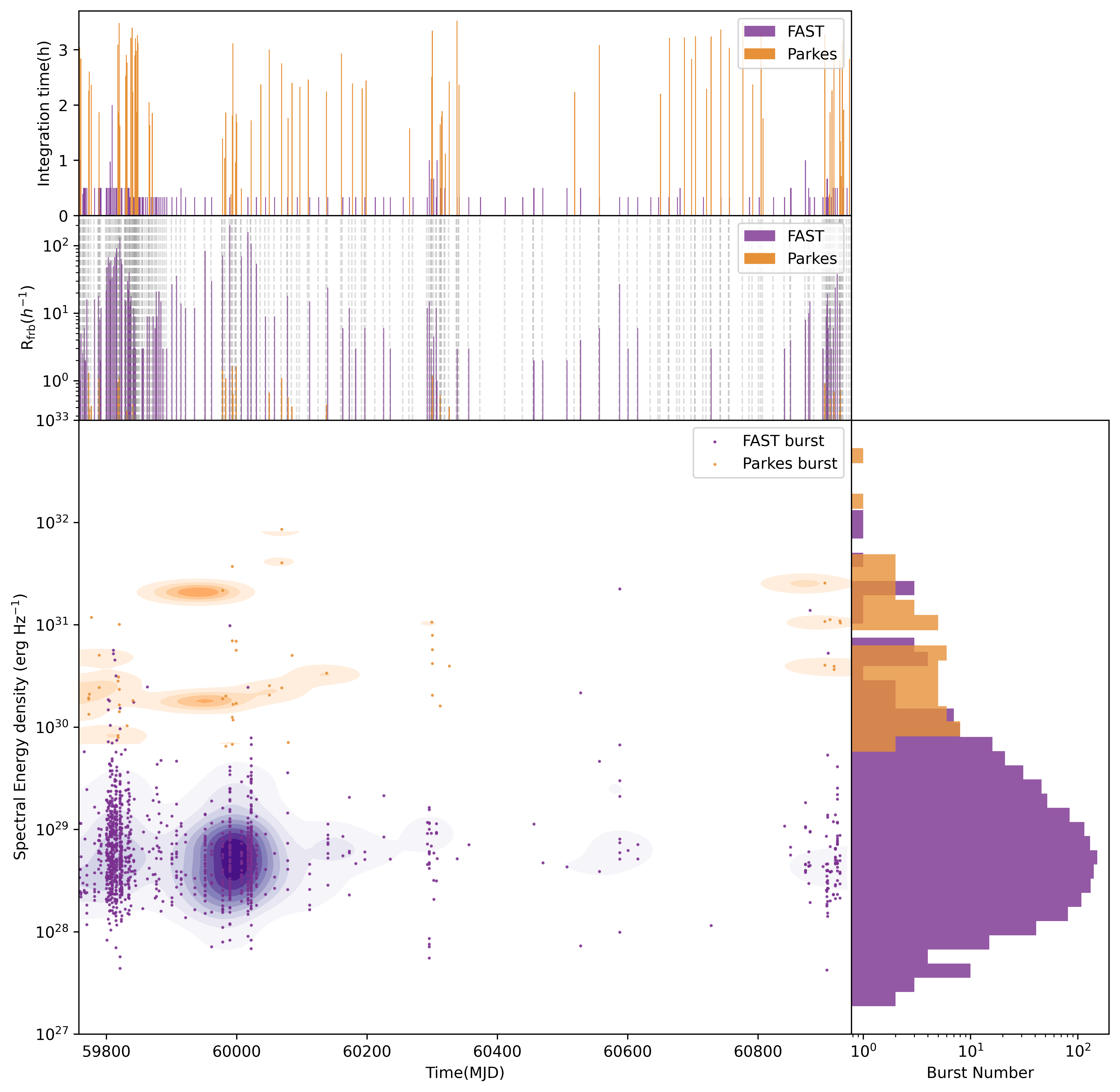}
        \caption{The figure shows the long-term temporal behavior and energetic diversity of FRB 20220529. (a) On-source integration time for FAST (purple) and Parkes (orange) observing sessions. (b) Evolution of the burst rate $R_{\mathrm{FRB}}$, revealing highly non-stationary activity superimposed on a secular fading trend. Grey dashed lines mark individual observing epochs. (c) Spectral energy density of individual bursts over the 3.2-year baseline. 
        %The overlaid density contours and the marginal histogram illustrate the vast dynamic range, with FAST capturing the low-energy bulk and Parkes resolving the rare, bright-end tail.  
        } 
        \label{figure:frb_obs}
	\end{figure*}

\clearpage

%% START OF METHODS 
\section*{Methods}

\setcounter{table}{0} % Reset figure counter
\captionsetup[table]{name={\bf Extended Data Table}}
\setcounter{figure}{0} % Reset figure counter
\captionsetup[figure]{name={\bf Extended Data Figure}}

\subsubsection*{FAST observational dataset}
Our observational campaign targeting FRB 20220529 commenced with two consecutive 1-hour grid mapping observations utilizing the 19-beam receiver on the FAST, starting at 23:13:00 Coordinated Universal Time~(UTC) on 22 June 2022.
Two individual bursts were detected in distinct receiver beams, facilitating an initial localization of the source at $\alpha = 01^{\rm h}16^{\rm m}23.35^{\rm s}$, $\delta = +20^\circ37^\prime34.7^{\prime\prime}$. 
Subsequent follow-up observations were performed using the on-source central beam of the receiver. 
Besides the on-source tracking, we carried out two additional grid mapping observations on 14 and 17 August 2022, complemented by an off-beam tracking observation on 28 August 2022. For these observations, bursts were detected simultaneously in up to three separate beams, enabling us to refine the source astrometry to $\alpha = 01^{\rm h}16^{\rm m}24.24^{\rm s}$, $\delta = +20^\circ38^\prime27.6^{\prime\prime}$. From 23 August 2022 through 8 September 2025, all observational runs were executed using the final astrometric solution refined via Karl G. Jansky Very Large Array (VLA) detections: $\alpha = 01^{\rm h}16^{\rm m}25.01^{\rm s}$, $\delta = +20^\circ37^\prime57^{\prime\prime}$. Our most recent tracking observation, conducted on 8 September 2025, resulted in the detection of 7 individual bursts from the source.
The 19-beam L-band receiver utilized for these observations covers a frequency passband of 1000–1500 MHz, split into 4096 spectral channels. Dual linear polarization signals were digitized with 8-bit sampling\cite{Jiang20RAA}, channelized using the Reconfigurable Open Architecture Computing Hardware generation 2 (ROACH 2) backend\cite{Hickish16}, and recorded in the PSRFITS search-mode data format\cite{Hotan04}. The time resolution of the raw data is 49.153 $\mu$s.
To perform flux and polarization calibration, a calibration signal with an equivalent noise temperature of 1 K was injected via a noise-switching scheme prior to the start of each observational run. The total on-source duration of each observation is presented in Fig.~\ref{figure:frb_obs}. The majority of our recent observations (up to and including 8 September 2025) were carried out in on-source tracking mode with a nominal duration of 0.5 hours; the latest observation yielded a measured burst rate of 14.0 bursts per hour. FRB rate estimates are unavailable for the earlier off-beam observations conducted on 22 June 2022 and 28 August 2022, while the burst rates for the 14 and 17 August 2022 grid observations were calculated using bursts detected during the first 0.5 hours of the run, when the source was within the primary beam of the central receiver.

\subsubsection*{Burst detection}
All observational data acquired with the FAST and Parkes radio telescopes were reduced and processed via two independent single-pulse search pipelines constructed using the standard pulsar and fast radio burst (FRB) search software packages \texttt{PRESTO}\cite{Ransom01} and \texttt{HEIMDALL}\cite{Petroff15a}.
For the FAST dataset, we performed the single-pulse search on the full instantaneous passband. For the Parkes UWL wideband data, we adopted a tiered sub-band search strategy splitting the data into a sequence of contiguous sub-bands spanning 128 to 3328 MHz \cite{Kumar21_11a}.
For both pipelines, the data were dedispersed over a DM interval of $200-300\,{\rm pc~cm^{-3}}$, with a uniform DM step size of 0.1$\,{\rm pc~cm^{-3}}$. Single-pulse candidates with a signal-to-noise ratio (S/N) above the 7$\sigma$ threshold were retained and subjected to manual visual vetting to exclude radio frequency interference (RFI) and spurious events.
In total, 1265 individual bursts were identified in the FAST observational dataset, of which 1215 bursts were detected during on-source pointing observations. This yields a mean on-source burst rate of 20.0 bursts per hour. For the Parkes monitoring dataset, we detected a total of 58 individual bursts. The burst activity light curve of FAST reveals two distinct outburst epochs centered on August 2022 and March 2023, reaching peak burst rates of 134 and 204 bursts per hour, respectively. When these periods of heightened activity are excluded, the mean quiescent burst rate of the source is 7.6 bursts per hour.
\subsubsection*{Spectral energy density calibration}
\label{sec:cal}
To estimate the burst flux densities $S$, we employed the radiometer equation:
\begin{equation}
S=\frac{{\rm S/N}\,T_{\rm sys}}{G \sqrt{{\Delta}{\nu}N_p{t}_{\rm obs}}},
\label{equ:limit}
\end{equation}
where we ignore the loss factor owing to the 8-bit sampling. S/N is the burst significance at the specified time resolution and frequency range of $t_{\rm obs} = 49.152\,\mu s$ and 1000$-$1500\,MHz (${\Delta}{\nu} = 500$\,MHz). For Parkes bursts, the corresponding $t_{\rm obs}$ and $\Delta\nu$ were taken from the UWL search setup used for each observation. $T_{\rm sys}$ is the system temperature: $\sim$23~K for the Parkes UWL receiver, and $\sim$20~K for the FAST 19-beam receiver. $G$ is the telescope antenna gain, with values of $\sim$ 0.757, and 16~K/Jy for the Parkes UWL, and FAST 19-beam receivers, respectively\cite{Jiang20RAA,uwlreceiver}. $N_p=2$ is the number of polarization channels.
The average burst fluence~($F$) is computed by integrating the burst flux with respect to time, and the equivalent width $W_{\rm eq}$ is derived by dividing the fluence by the burst peak flux. The sample completeness was determined with the following method. We simulated 10,000 mock bursts with Gaussian temporal profile and bandpass matching the detected distributions. We then randomly injected the mock bursts into the original FAST data when no FRB was detected. The mock-burst-injected data are then fed to our burst-searching pipeline to compute the detection rate. The procedure shows that the fluence threshold that achieves the 95\% detection probability with $S/N\ge 7$ is 22.6 mJy ms and 398.1 mJy ms for the FAST 19-beam receiver and Parkes UWL, respectively. The spectral energy density $E_\nu$ was calculated by integrating over the $4\pi$ solid angle, $E_\nu={4\pi D_L^2}(1+z)^{-1}\,F$, where $D_L =923.2\rm Mpc$  is the luminosity distance. 

\subsubsection*{Statistical fitting and robustness analysis }

All statistical analyses and core fitting results presented in this work are based on the on-source burst sample from the FAST telescope to mitigate systematic biases arising from cross-instrument calibration and heterogeneous observing strategies. The Parkes dataset is reserved for independent verification of the high-energy power-law tail and was not included in the core fitting and robustness tests described below. Comprehensive details of the fitting procedures, convergence tests, and robustness checks are provided in the following paragraph.

\paragraph{1. Fitting model definitions}
We model the cumulative burst-rate distribution $R(>E)$, defined as the number of bursts detected per hour with spectral energy density greater than $E$, as the primary observable for all fits. 

The core two-component exponential plus power-law (EXP+PL) model, which provides the statistically preferred description of the data, takes the form:
\begin{equation}
R(>E) = A \exp\left(-\frac{E}{E_0}\right) + B \left( \frac{E}{10^{30}\ \mathrm{erg\ Hz^{-1}}} \right)^{-\gamma}
\end{equation}
where $A$ and $B$ (units: bursts~h$^{-1}$) are the amplitude factors for the exponential and power-law components, respectively, $E_0$ (units: $\mathrm{erg\ Hz^{-1}}$) is the characteristic energy scale of the exponential component and $\gamma$ is the power-law index of the high-energy tail. The $10^{30}\ \mathrm{erg\ Hz^{-1}}$ term is a fixed pivot energy scale adopted solely to nondimensionalize the energy argument.

We define the crossover energy $E_{\nu,c}$ as the energy where the exponential and power-law components contribute equally to the cumulative burst rate, satisfying the equality:
\begin{equation}
A \exp\left(-\frac{E_{\nu,c}}{E_0}\right) = B \left( \frac{E_{\nu,c}}{10^{30}\ \mathrm{erg\ Hz^{-1}}} \right)^{-\gamma}
\end{equation}

To rigorously validate the statistical superiority of the EXP+PL model, we conducted head-to-head comparisons against five alternative models:

1.  Single power-law (PL) model: $R(>E) = B \left( \frac{E}{10^{30}\ \mathrm{erg\ Hz^{-1}}} \right)^{-\gamma}.$

2.  Single exponential (EXP) model: $R(>E) = A \exp\left(-\frac{E}{E_0}\right).$

3.  Broken power-law (BPL) model:
\begin{equation}
R(>E) =
\begin{cases}
B \, \left( \frac{E}{E_b}\right)^{-\gamma_1}, & E \leq E_b \\[8pt]
B \, \left( \frac{E}{E_b}\right)^{-\gamma_2}, & E > E_b
\end{cases} \nonumber
\end{equation}

4.  Smoothly broken power-law (SBPL) model:
\begin{equation}
R(>E) = B \left( \frac{E}{E_b} \right)^{-\alpha_1} \left[ 1 + \left( \frac{E}{E_b} \right)^{1/s} \right]^{(\alpha_1 - \alpha_2) s}
\label{eq:sbpl}
\end{equation}

5.  Double exponential (DEXP) model: $R(>E) = A_1 \exp\left(-\frac{E}{E_{0,1}}\right) + A_2 \exp\left(-\frac{E}{E_{0,2}}\right).$

\begin{figure*}
\centering
\includegraphics[width=0.7\textwidth]{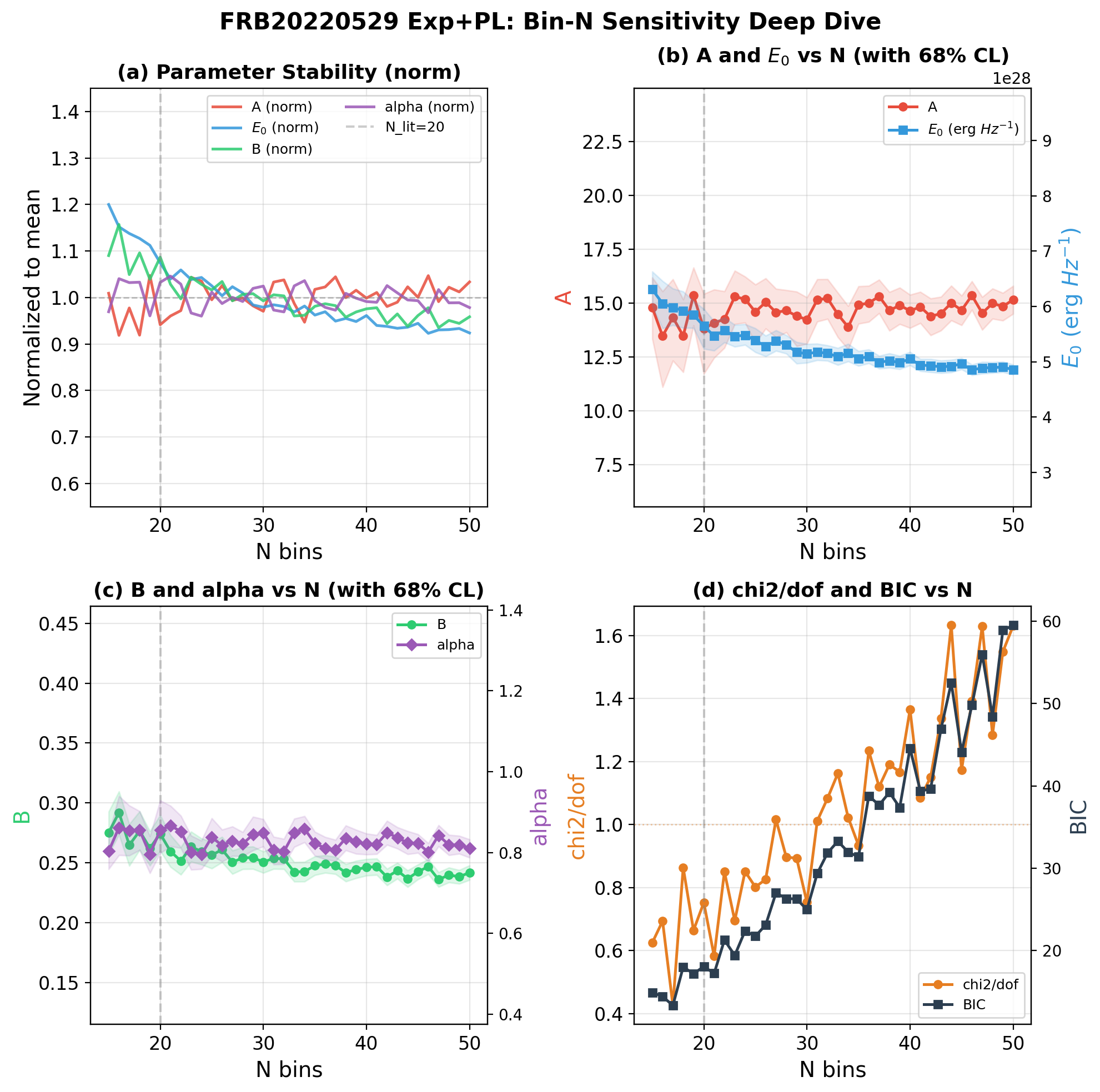}
\caption{\textbf{Robustness of the fitting results against the choice of bin number.} The best-fit parameters of the EXP+PL model remain stable across bin numbers from $N=15$ to $N=50$, demonstrating that the adopted binning prescription ($N=20$) does not affect the conclusions.}
\label{fig:C3}
\end{figure*}

All fits were restricted to the energy range above the 95\% completeness threshold of the FAST sample to minimize detection-limit induced biases. For all fits, we adopt logarithmically uniform energy bins. The number of bins is chosen following the empirical prescription $N = \log_{10}(E_{\max}/E_{\min}) \times 5$~\cite{Aschwanden_2015}, yielding $N=20$ for the FAST sample. We require that the burst count in each bin differs from adjacent bins by at least one event, thereby avoiding duplicated cumulative points in the fitted distribution. To verify that the choice of bin number does not affect our conclusions, we tested a wide range from $N=15$ to $N=50$ and found that the model selection results and best-fit parameters remain consistent across this range (Extended Data Fig.~\ref{fig:C3}).

\paragraph{2. Fitting algorithm and uncertainty quantification}
We employ a Bayesian Markov Chain Monte Carlo (MCMC) framework via the Python \texttt{emcee} affine-invariant ensemble sampler to explore the parameter space and quantify posterior uncertainties. This approach accounts for the Poisson nature of burst count statistics and avoids biases associated with least-squares fitting in low-count regimes.

For all models, we adopt weakly informative prior distributions: normalization parameters $A, B$ follow a uniform prior over $[0.01, 1000]\ \mathrm{bursts\ h^{-1}}$; Characteristic energy $E_0$ follows a log-uniform prior over $[10^{28}, 10^{31}]\ \mathrm{erg\ Hz^{-1}}$ and the power-law index $\gamma$ follows a uniform prior over $[0, 10]$. For the alternative empirical models, we used analogous weakly informative priors over physically allowed parameter ranges.

We run the MCMC sampler with 64 walkers, a burn-in phase of 8000 steps, and a production phase of 20000 steps. We report the median of the posterior distribution as the best-fit value for each parameter, with uncertainties defined by the 95\% credible interval.

\paragraph{3. Model comparison criteria}
To quantitatively evaluate the performance of competing models, we utilize two complementary statistical metrics: the reduced chi-squared statistic ($\chi^2/\mathrm{dof}$) and the Bayesian Information Criterion (BIC).

The reduced chi-squared statistic is calculated as:
\begin{equation}
\chi^2/\mathrm{dof} = \frac{1}{N - k} \sum_{i=1}^{N} \frac{\left( R_{\mathrm{obs},i} - R_{\mathrm{mod},i} \right)^2}{\sigma_{\mathrm{obs},i}^2}
\end{equation}
where $N$ is the number of energy bins, $k$ is the number of free parameters in the model, $R_{\mathrm{obs},i}$ and $R_{\mathrm{mod},i}$ are the observed and model-predicted cumulative burst rates in bin $i$, respectively, and $\sigma_{\mathrm{obs},i} = R_{\mathrm{obs},i} / \sqrt{N_{\mathrm{burst},i}}$ is the Poisson uncertainty on the observed rate, with $N_{\mathrm{burst},i}$ representing the number of bursts in bin $i$. A value of $\chi^2/\mathrm{dof} \approx 1$ indicates a statistically acceptable fit.

The BIC is calculated to penalize over-parameterization and quantify the relative strength of evidence for each model:
\begin{equation}
\mathrm{BIC} = -2 \ln(\mathcal{L}_{\mathrm{max}}) + k \ln(N)
\end{equation}
where $\mathcal{L}_{\mathrm{max}}$ is the maximum likelihood value of the model. A difference in BIC ($\Delta\mathrm{BIC}$) of $>10$ between two models indicates decisive statistical evidence in favor of the model with the lower BIC value. Because neighboring cumulative-rate points are not strictly independent, the reduced $\chi^2$ statistic is used primarily as a goodness-of-fit diagnostic, whereas BIC is used for relative model comparison under the same binning and likelihood prescription.

\begin{figure*}
		\centering

    \includegraphics[width=150mm]{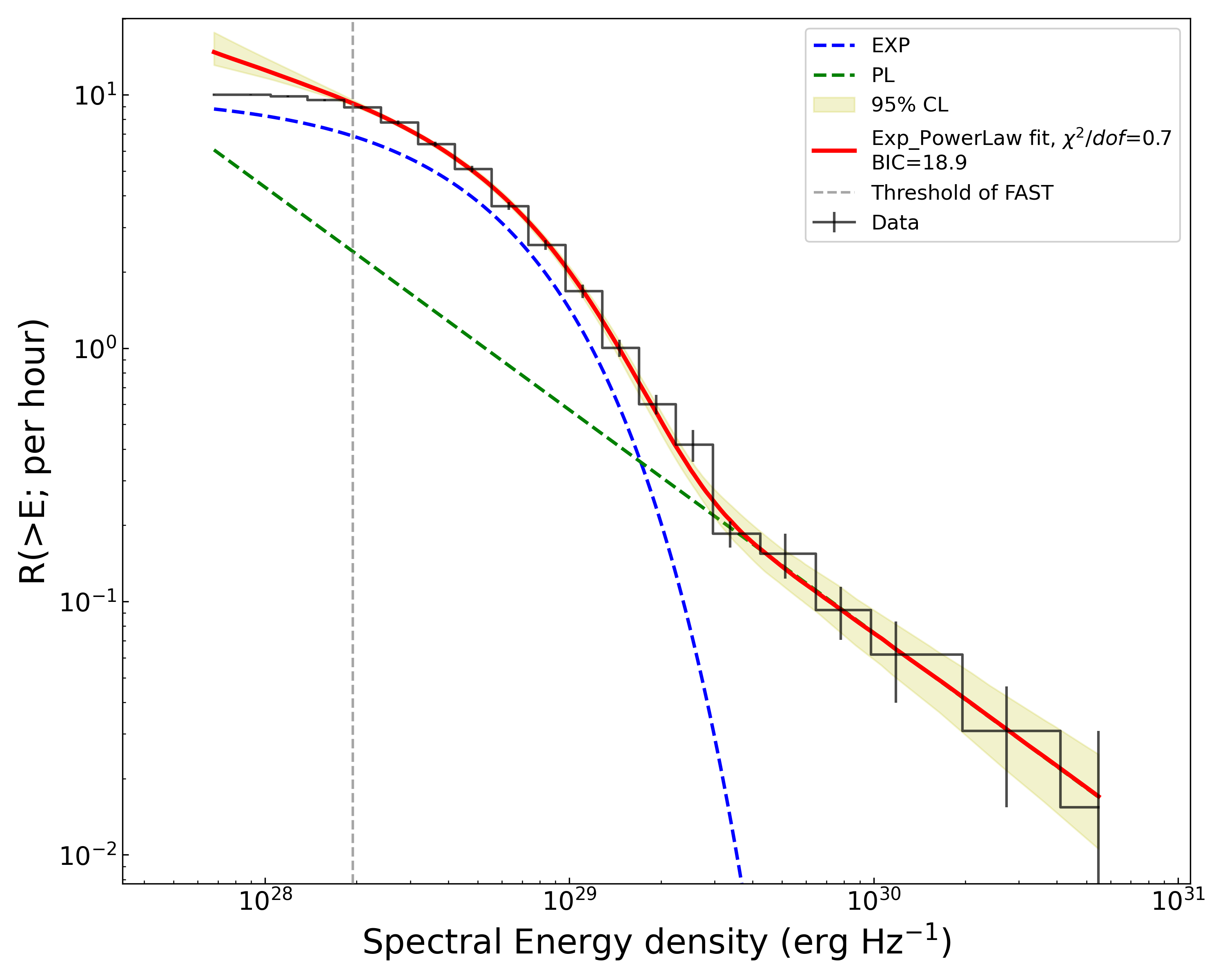}

		\caption{The cumulative burst-rate distributions of FRB 20220529 full passband observation bursts.}
		\label{fig:fullband}
	\end{figure*}

\paragraph{4. Epoch-resolved analysis and temporal stability tests}
To assess the long-term temporal evolution of the energy distribution, we partitioned the full FAST on-source sample into four chronological epochs, ordered by the detection time of individual bursts. Each epoch contains approximately 300 bursts, ensuring comparable sample size and uniform statistical power across all epochs, and reducing biases from non-stationary count statistics. The epoch boundaries are defined as follows:
\begin{enumerate}
\item Epoch 1: MJD 59758 – 59812 (303 bursts),
\item Epoch 2: MJD 59812 – 59835 (303 bursts),
\item Epoch 3: MJD 59835 – 60022 (303 bursts),
\item Epoch 4: MJD 60022 – 60925 (306 bursts),
\end{enumerate}
We fit the EXP+PL model to each epoch independently, using the identical Bayesian framework and completeness threshold applied to the full sample. To verify that the observed temporal stability, particularly that of the characteristic scale $E_0$, is not an artifact of the specific epoch definition or of the treatment of the high-energy tail, we conducted four robustness checks:
\begin{enumerate}

\item Alternative splitting schemes: We tested equal-time epoch splits and activity-based splits (separating enhanced-activity and quiescent periods), and found consistent behaviour of the characteristic scale across all splitting schemes;
\item Rolling window analysis: We applied the EXP+PL model to a 300-burst rolling window stepped by 50 bursts across the full dataset, confirming that $E_0$ remains consistent within the 95\% credible intervals across the entire 3.2-year baseline.
\item Fixed-$\gamma$ consistency check: To verify that the central finding of $E_0$ invariance is not sensitive to epoch-to-epoch variations in the fitted power-law index, we performed fixed-$\gamma$ fits for each epoch using the full-sample value $\gamma = 0.85$. The resulting $E_0$ values remain consistent within the quoted uncertainties across all epochs, with Epoch~4 yielding $E_0=6.07\times10^{28}\ {\rm erg\ Hz^{-1}}$, supporting the interpretation that the apparent late-time flattening of the high-energy tail does not affect the stability of the exponential scale.
\item Full-bandwidth sample consistency check: The FAST receiver covers a bandwidth of 500~MHz (1000~MHz to 1500~MHz). Some bursts may be affected by the limited bandwidth of the receiver, where only a fraction of the full passband contributes to the detection. To verify that the EXP+PL model preference and the stability of $E_0$ are not biased by such effects, we restricted the analysis to the 651 bursts for which the full burst bandwidth was contained within the FAST passband. Refitting all models to this full-bandwidth sub-sample yields results fully consistent with the full-sample inference: the EXP+PL model remains preferred, and the characteristic energy scale, \(E_0 = 5.14 \times 10^{28}\ {\rm erg\,Hz^{-1}}\), remains consistent with the full-sample value~(Extended Data Table~\ref{table:energy_fit_fullband} and Extended Data Fig.~\ref{fig:fullband}). This confirms that limited bandwidth does not drive our conclusions.
\end{enumerate}

	\begin{table*}[htbp]
		\caption{\bf Full passband observation burst energy ranges and fitting results of energy distribution for FRB 20220529.}
		\renewcommand\arraystretch{1.1}
		\begin{center}
			\begin{threeparttable}
				\begin{tabular}{lcc}
					\hline
					{\bf Property} & \multicolumn{2}{c}{\bf Measurement} \\
					\hline
					% Energy ranges (单独占一行跨列显示)
					FAST energy density~(erg~Hz$^{-1}$) & \multicolumn{2}{c}{$5.9\times10^{27} - 6.3\times10^{30}$} \\
					\hline
					{\bf Model} & $\mathbf{\chi^{2}/dof}$ & {\bf BIC} \\
					\hline
					
					Single power law & 30.2 & 428.9 \\
					Single exponential & 11.9 & 171.5 \\
					[0.5em]
					Broken power law & 2.2 & 37.7 \\
                    Smoothly broken power law & 2.1 & 36.5 \\
					Double exponential & 1.6 & 30.2 \\
					Exponential + power law & 0.7 & 18.9 \\
					\hline
				\end{tabular}
				%   \begin{tablenotes}
					%        \footnotesize
					%        \item[a]Corrected to .
					
					%     \end{tablenotes}
			\end{threeparttable}
		\end{center}
		\label{table:energy_fit_fullband}
	\end{table*}

\paragraph{5. Scale-free robustness tests}

To further test whether the high-energy component of FRB~20220529 exhibits scale-free, avalanche-like statistics, we applied a canonical $q$-Gaussian distribution to fit the return distribution ($x_n$) of the full dataset and specific high-energy subsets ($>2\times10^{29}\,{\rm erg\,Hz^{-1}}$ and $>5\times10^{29}\,{\rm erg\,Hz^{-1}}$). In statistical mechanics, when a process follows a scale-invariant power law, the fitted $q$ value remains invariant with respect to the sampling scale $n$ \cite{2015EPJB...88..206W}. Our results demonstrate that the full dataset exhibits a clear lack of scale invariance, consistent with the presence of a characteristic energy scale in the dominant low-energy population. In contrast, the high-energy subsets exhibit a pronounced convergence toward scale invariance. Notably, in the highest-energy regime, where contamination from low-energy stochastic components is weakest, the derived $q$ value closely follows the theoretical relation linking the power-law index $\alpha$ to the entropic index $q$: $\alpha = 2/(q-1) = 2/(2.17-1) \approx  1.71$  \cite{2010PhRvE..82b1124C}. This independent statistical diagnostic provides a consistency check that the high-energy subset behaves more nearly scale-invariant than the full sample.

The Tsallis $q$-Gaussian distribution provides a framework for quantifying deviations from Gaussian statistics ($q=1$). We examine the distribution of $X_n$, defined here in terms of energy fluctuations. Let $S_i$ be the energy of the $i$th event; then we define
\[
X_n = S_{i+n} - S_i,
\]
where $n$ is the event-index interval. To reduce computation, we rescale $X_n$ by its standard deviation $\sigma_{X_n}$ as
\[
x_n = X_n / \sigma_{X_n}.
\]
The differential distribution of $x_n$ is described by \cite{1998PhyA..261..534T},
\begin{equation}
f(x_n) = A \bigl[ 1 - B(1-q) x_n^2 \bigr]^{1/(1-q)}.
\end{equation}
To account for the limited sample size in the high-energy regime, we performed the fitting using the cumulative distribution function (CDF) rather than the differential form. The CDF is defined as \cite{Wei2023PhRvR}
\begin{equation}
F(x_n) = \int_{-\infty}^{x_n} f(x')\,dx'.
\end{equation}

As noted above, a theoretical relationship exists between the power-law index $\alpha$ of the energy distribution and the $q$ value \cite{2010PhRvE..82b1124C}:
\begin{equation}
\alpha = \frac{2}{q-1}.
\label{eq:alpha_q}
\end{equation}
This relation allows a consistency check between the $q$-Gaussian analysis and the high-energy power-law behaviour inferred from the primary rate-distribution fits.

	\begin{figure*}
		\centering
		\includegraphics[width=50mm]{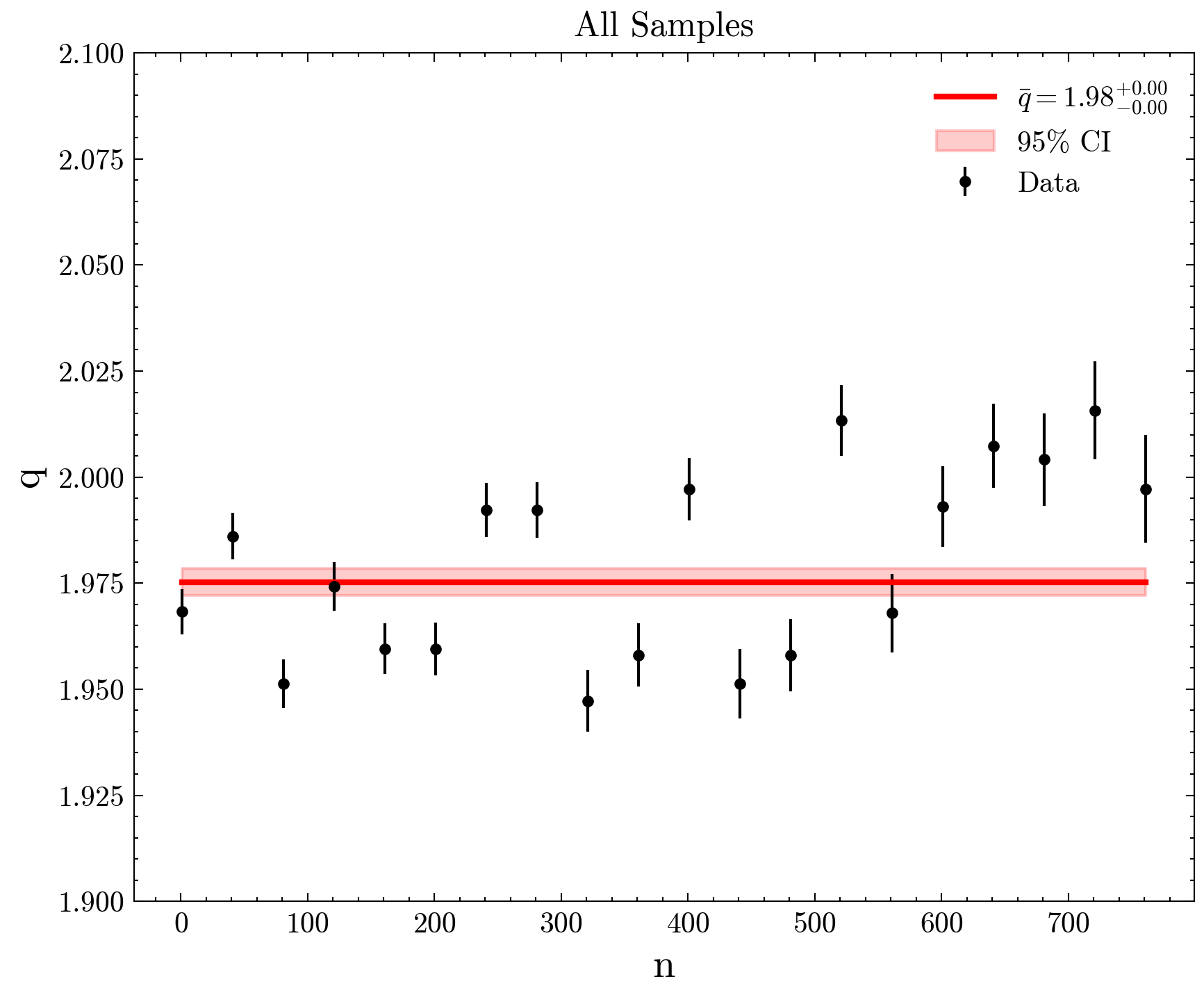}
        \includegraphics[width=50mm]{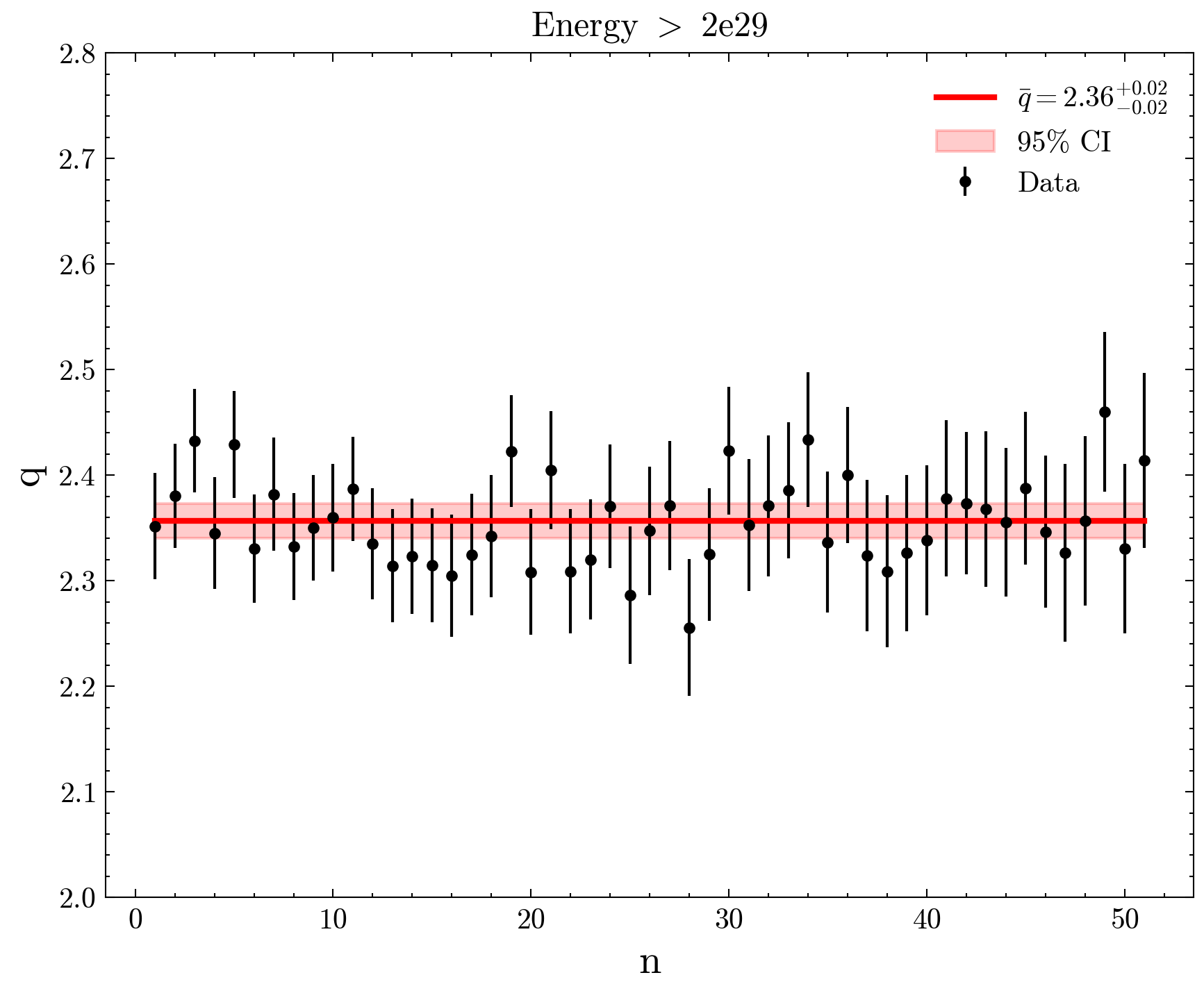}
        \includegraphics[width=50mm]{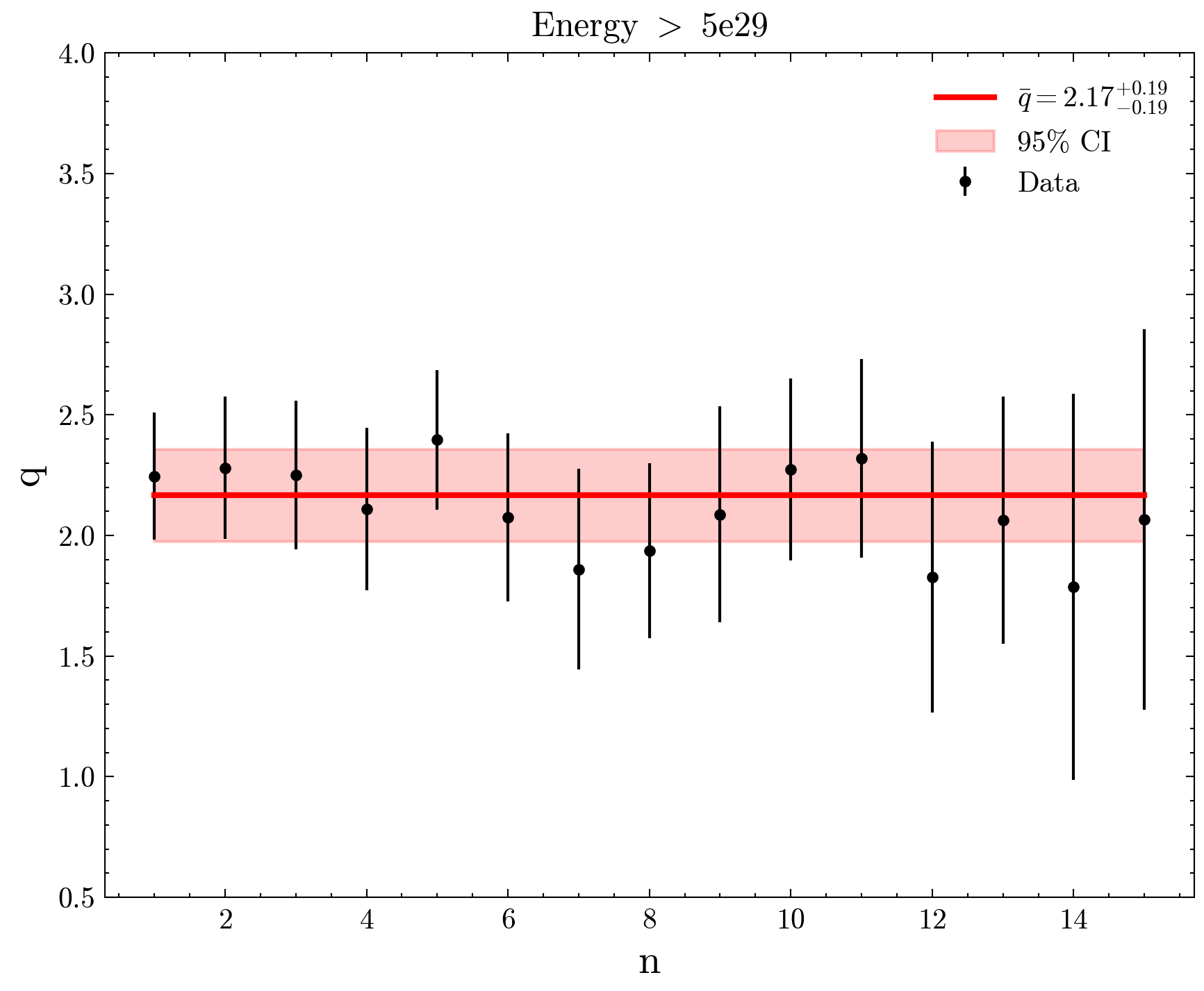}
		\caption{Tsallis $q$-Gaussian analysis of the burst-energy return distribution for FRB 20220529. (a) Full energy sample; (b) bursts with $E_\nu>2\times10^{29}$ erg Hz$^{-1}$; (c) bursts with $E_\nu>5\times10^{29}$ erg Hz$^{-1}$. The high-energy subsets show more nearly scale-invariant behaviour than the full sample.}
		\label{fig:q-gauss}
	\end{figure*}
    
\subsubsection*{High-energy tail verification}
The Parkes ultra-wideband dataset presented in this work is reserved exclusively for independent verification of the high-energy power-law tail identified in the FAST-only core statistical fit. This cross-validation is designed to test whether the scale-free high-energy component is robust against statistical fluctuations, sample incompleteness and instrumental systematics.

We performed long-term monitoring observations of FRB 20220529 using the Parkes 64-m radio telescope, equipped with the Ultra-Wideband Low (UWL) receiver, covering the period from 27 June 2022 to 8 September 2025. The pointing strategy for all Parkes observations was matched to that of the contemporaneous FAST observational campaign to ensure consistent on-source coverage. 
The UWL receiver system provides an instantaneous frequency coverage of 704–4032 MHz\cite{uwlreceiver}. Data were recorded with 2-bit digitization, with two standard setups adopted throughout the campaign: a time resolution of 32 $\mu$s paired with 1 MHz-wide spectral channels, and a time resolution of 256 $\mu$s paired with 0.125 MHz-wide spectral channels.
Prior to 27 September 2022, the data were coherently de-dispersed at a fiducial dispersion measure (DM) of 247$\,{\rm pc~cm^{-3}}$, with only a single recorded polarization channel. Full Stokes parameters have been acquired for all observations conducted after this date. Our most recent Parkes observation, performed on 8 September 2025, had a total on-source duration of 1.34 hours and resulted in the detection of 1 burst.
For polarization and flux calibration, a 2-minute noise diode signal was injected into the signal chain prior to each tracking observation. 
In total, we acquired 81 Parkes observing sessions with a total on-source integration time of 204.2 hours, detecting 58 high-confidence bursts from FRB 20220529, with spectral energy densities ranging from $6.5\times10^{29}$ to $8.6\times10^{31}\ \mathrm{erg\ Hz^{-1}}$.

We performed independent flux and spectral energy density calibration for the Parkes dataset. For each detected burst, we calculated the peak flux density via the calibrated PSRFITS files with PSRCHIVE software. The system equivalent flux density (SEFD) of the noise source was calibrated using measurements from the Parkes Pulsar Timing Array project. The CAL file was obtained by making an observation pointing directly towards a source with known flux density (B0407-658) with the noise source switching. Following flux calibration with corresponding calibration files and unit conversion to janskys~(Jy), we directly derived the flux density of individual pulses from the calibrated files. The burst fluence~($F$) and spectral energy density were then computed using the same process applied to the FAST data.

We performed an independent power-law fit to verify that the Parkes high-energy bursts are consistent with an extended bright-end tail, using the identical MCMC fitting framework and priors applied to the FAST core dataset. The Parkes-only fit gives a power-law index of $\gamma=0.64\pm0.02$, confirming that the Parkes bursts follow an extended power-law-like bright-end distribution. Given the different observing band, sensitivity and limited Parkes sample size, we use this result as an external validation of the presence of a scale-free high-energy tail rather than as a precision measurement of the same slope.

	\begin{figure*}
		\centering
    \begin{tabular}{cc}
    \includegraphics[width=75mm]{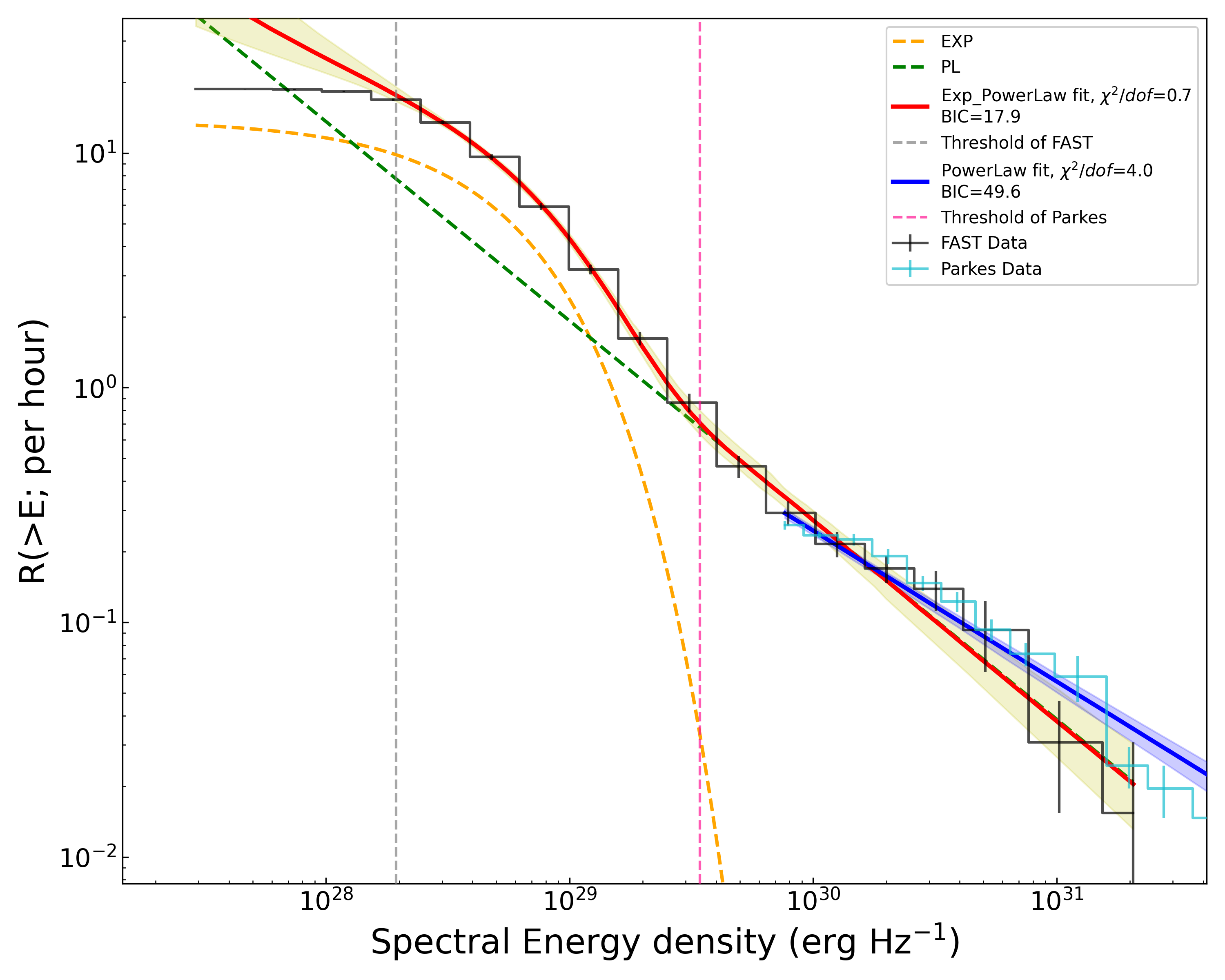} & 
    \includegraphics[width=75mm]{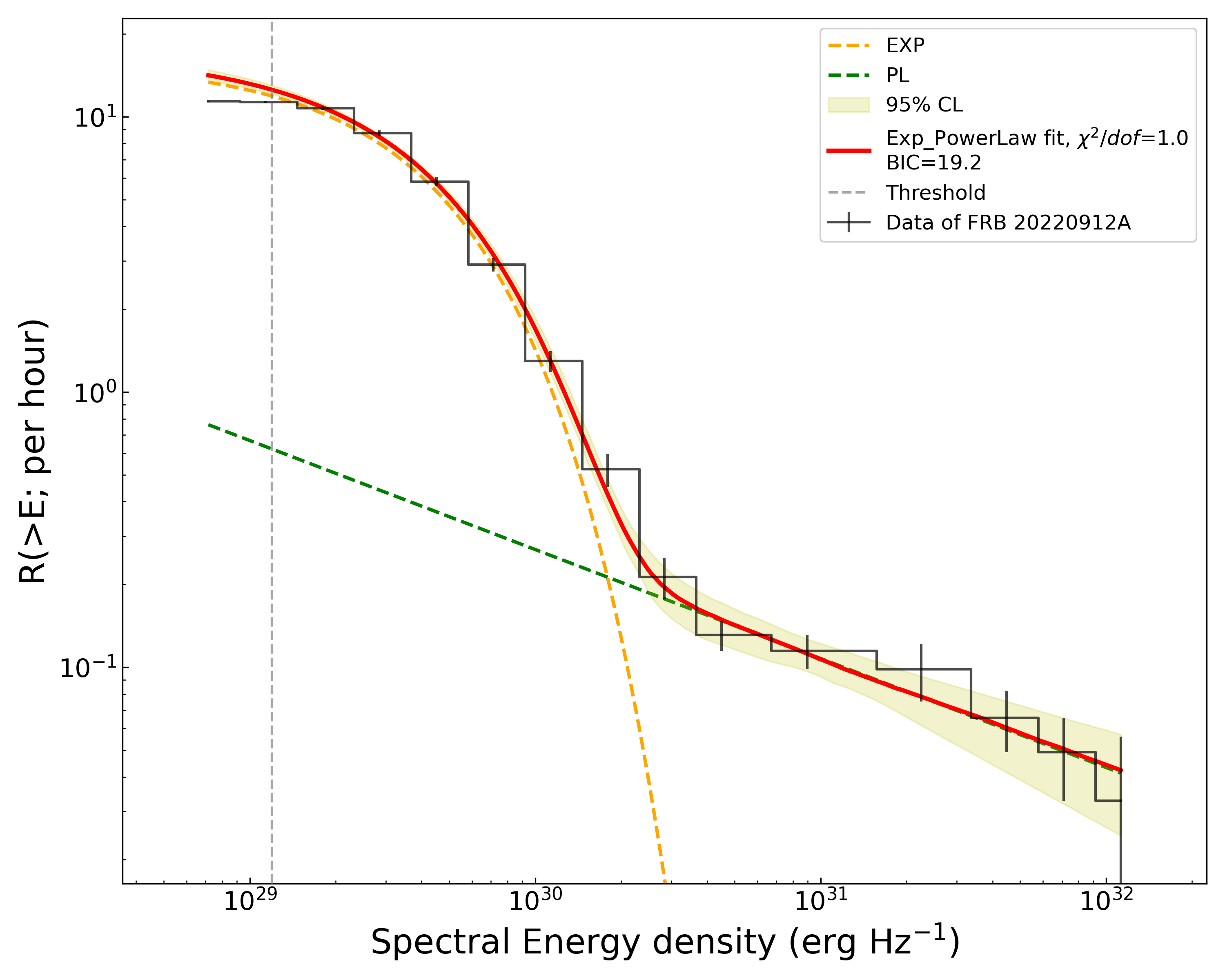}\\
    \end{tabular}
		\caption{High-energy tail verification and comparison with FRB~20220912A. (Left) Joint distribution for FRB 20220529, where Parkes detections confirm a power-law continuation of the high-energy tail. (Right) Observed cumulative burst-rate distribution for the hyperactive repeater FRB~20220912A, displaying a similar exponential-to-power-law transition. The phenomenological alignment between these two sources suggests an analogous two-component energy hierarchy.}
		\label{fig:compare0912}
	\end{figure*}

\section*{Comparison with other hyperactive repeating FRBs}
We performed a comparative analysis of four other well-characterized hyperactive repeaters: FRB 20121102, FRB 20201124A, FRB 20220912A, and FRB 20240114A. We applied the same cumulative burst-rate fitting framework, Bayesian inference procedure, model-comparison criteria and completeness-threshold criterion used for FRB 20220529 to these sources. 

Among these sources, FRB~20220912A provides the clearest independent comparison, with 696 bursts spanning approximately four orders of magnitude in energy from long-term single-telescope monitoring~\cite{Konijn2024}. 
For this source, the EXP+PL model yields a statistically preferred fit with best-fit parameters $E_0 = 4.2\times10^{29}\ \mathrm{erg\ Hz^{-1}}$, $\gamma = 0.40$. The EXP+PL model is preferred over the tested alternatives~(Extended Data Table~\ref{table:energy_frb0912}). For FRB 20220529, the inclusion of Parkes detections is important because it shows that the high-energy tail remains consistent with an extended power-law continuation rather than an exponential cutoff. The comparison with FRB 20220912A therefore does not rely solely on a qualitative resemblance, but on the shared presence of a characteristic-scale component together with a bright-end tail.

For FRB 20121102, FRB 20201124A, and FRB 20240114A, we find no significant preference for the EXP+PL model. This null result could be driven by non-uniform sampling, limited energy dynamic range or intrinsic temporal clustering of burst events. Long-term, uniformly cadenced monitoring of these sources will test whether the dual-mode signature emerges once their activity cycles are more completely sampled.

We therefore treat these sources as currently inconclusive for the presence of a resolved EXP+PL structure, rather than as population-level counterexamples. 
Long-term, high-cadence and wide-band monitoring will be required to determine how common such two-component energy hierarchies are among active repeating FRBs. A detailed discussion of how observational selection effects and intrinsic source evolution influence these null results is provided in Supplementary Note 5.

	\begin{table*}[htbp]
		\caption{\bf Energy ranges and fitting results of energy distribution of FRB 20220912A.}
		\renewcommand\arraystretch{1.1}
		\begin{center}
			\begin{threeparttable}
				\begin{tabular}{lcc}
					\hline
					{\bf Property} & \multicolumn{2}{c}{\bf Measurement} \\
					\hline
					% Energy ranges (单独占一行跨列显示)
					FRB 20220912A energy density~(erg~Hz$^{-1}$) & \multicolumn{2}{c}{$5.7\times10^{28} - 1.4\times10^{32}$} \\
					\hline
					{\bf Model} & $\mathbf{\chi^{2}/dof}$ & {\bf BIC} \\
					\hline
					
					Single power law & 28.9 & 323.3 \\
					Single exponential & 18.0 & 203.6 \\
					[0.5em]
					Broken power law & 7.2 & 75.3 \\
                    Smoothly broken power law & 8.2 & 78.3 \\
					Double exponential & 1.3 & 21.9 \\
					Exponential + power law & 1.0 & 19.2 \\
					\hline
				\end{tabular}
				%   \begin{tablenotes}
					%        \footnotesize
					%        \item[a]Corrected to .
					
					%     \end{tablenotes}
			\end{threeparttable}
		\end{center}
		\label{table:energy_frb0912}
	\end{table*}

\section*{Supplementary Note 1. Temporal evolution of the fit parameters}\label{supsec:temp_evo}

The epoch-resolved fits in Extended Data Table~\ref{tab:4epoch_fit} show that the temporal evolution is dominated by changes in the component normalizations rather than by a drift of the characteristic energy scale. Across all four epochs, \(E_0\) remains confined within \((4.83\)--\(6.82)\times10^{28}\,{\rm erg\ Hz^{-1}}\). By contrast, the amplitudes \(A\) and \(B\) vary with activity level and are substantially reduced in the late-time, low-activity epoch. This suggests that the source fades mainly by producing fewer burst-triggering dissipation events, rather than by shifting the local energy scale of the low-energy component.

The high-energy index \(\gamma\) is also broadly stable during the burst-rich phase. In Epochs 1--3, \(\gamma\) remains within \(0.74\)--\(0.99\), indicating that the bright-end cascade component is approximately preserved. Epoch 4 gives a substantially flatter value, \(\gamma=0.39\), but this late-time interval contains fewer bright bursts and is more susceptible to small-number statistics at the high-energy end. We therefore adopt the full-sample fit as the fiducial constraint, while using the epoch-resolved fits as a robustness check.

Overall, the time-resolved analysis supports a stable dissipation hierarchy during the statistically well-sampled portion of the monitoring campaign: the localized component preserves a nearly invariant \(E_0\), and the high-energy tail remains consistent in Epochs 1--3. The late-time deviation is interpreted conservatively as a consequence of reduced statistics in the low-activity regime.

\section*{Supplementary Note 2. Mathematical basis of the thresholded dissipation hierarchy}\label{supsec:mathbas}

The observed exponential-plus-power-law (EXP+PL) energy distribution is interpreted within the framework of multi-scale magnetic reconnection in a twisted magnetar magnetosphere. In this picture, magnetic stresses build up in the closed-field region and intermittently generate elongated current sheets \cite{Thompson2002,Beloborodov2009,Parfrey2012,Parfrey2013}. The resulting burst energetics are governed not only by whether plasmoids can form, but also by how far a local reconnection trigger can propagate into a collective cascade \cite{Philippov2019,Lyubarsky2020,Beloborodov2020,Yuan2020,Mahlmann2022,Wang2023,Burnaz2025,Long2025,Gourgouliatos2026}.

We distinguish the global Lundquist number of the current sheet from the effective branching state of an individual reconnection episode. We denote the Lundquist number by \(S_{\rm L}\), to avoid confusion with the avalanche size introduced below. The condition \(S_{\rm L}>S_c\) allows plasmoid formation, but it does not require every local trigger to develop into an observable macroscopic avalanche. The EXP+PL morphology is therefore interpreted as the statistical coexistence of localized subcritical reconnection episodes and near-critical cascade episodes within the same magnetospheric environment. The near-critical branch can involve fragmentation, interaction and merger of plasmoids and secondary current sheets, producing scale-free dissipation consistent with the observed high-energy tail \cite{Samtaney2009,Uzdensky2010,HuangBhattacharjee2010,HuangBhattacharjee2012,LoureiroUzdensky2016}. 

\paragraph{Global Lundquist number in the magnetosphere.}

For typical magnetar parameters, the global Lundquist number is expected to be very large. As an order-of-magnitude reference for the scale hierarchy, we first estimate the collisional Spitzer resistivity. The Spitzer electric resistivity of the plasma is
\begin{equation}
	\rho_{\rm Sp}=
	\frac{\pi e^2m_e^{1/2}\ln\Lambda_{\rm C}}
	{(k_BT_e)^{3/2}},
\end{equation}
where the Coulomb logarithm \(\ln\Lambda_{\rm C}\) is
\begin{equation}
	\ln\Lambda_{\rm C}=
	\ln\left[
	\frac{3k_B^{3/2}}{2\pi^{1/2}e^3}
	\left(\frac{T_e^3}{n_e}\right)^{1/2}
	\right],
\end{equation}
and \(T_e\), \(n_e\) are the electron temperature and number density within the magnetosphere, respectively. The local dipole field at an emission radius \(R_{\rm em}=10^8r_8\,{\rm cm}\) is
\begin{equation}
	B_{\rm loc}\simeq B_s\left(\frac{R_*}{R_{\rm em}}\right)^3
	\simeq
	10^9B_{{\rm s},15}R_6^3r_8^{-3}\ {\rm G}.
	\label{eq:B_loc}
\end{equation}
The Goldreich--Julian density is then
\begin{equation}
	n_{\rm GJ}\simeq
	\frac{\Omega B_{\rm loc}}{2\pi ec}
	=
	\frac{B_{\rm loc}}{ecP}
	\simeq
	6.94\times10^7
	B_{{\rm s},15}R_6^3P_{\rm s}^{-1}r_8^{-3}\,
	{\rm cm^{-3}},
\end{equation}
where \(P_{\rm s}\) is the spin period in seconds. Substituting \(n_e=\mathcal{M}n_{\rm GJ}\) with a typical pair multiplicity \(\mathcal{M}=10^3\mathcal{M}_3\), and adopting \(T_e=10^4\,{\rm K}\), gives
\begin{equation}
	\rho_{\rm Sp}\simeq 0.14\ \Omega\,{\rm cm},
\end{equation}
and the corresponding magnetic diffusivity in cgs units is
\begin{equation}
	\eta_m\simeq 1.15\times10^7\ {\rm cm^2\,s^{-1}} .
\end{equation}
Given that the magnetization remains very high even for large pair multiplicities, we take \(v_A\simeq c\). For a macroscopic current sheet of length \(L\sim R_{\rm em}=10^8\,{\rm cm}\), the Lundquist number is then \cite{Samtaney2009}
\begin{equation}
	S_{\rm L}=\frac{Lv_A}{\eta_m}
	\simeq 2.6\times10^{11},
\end{equation}
far above the critical plasmoid threshold \(S_c\sim10^4\). Equivalently, the length required to reach \(S_{\rm L}=S_c\) is only
\begin{equation}
	L_c=\frac{S_c\eta_m}{c}\simeq 3.8\ {\rm cm}.
\end{equation}

This estimate is intended only to establish the scale hierarchy. A macroscopic magnetospheric current sheet should already lie deeply in the plasmoid-unstable regime. The low-energy exponential component therefore should not be read as evidence for globally stable \(S_{\rm L}<S_c\) current sheets; it instead indicates that many local plasmoid-forming triggers remain confined rather than propagating into extended avalanches. Non-ideal, collisionless or anomalous resistive effects may change the numerical diffusivity, but they do not alter the qualitative conclusion that macroscopic magnetospheric sheets are far above the nominal plasmoid threshold.

\paragraph{Observed burst as a reconnection avalanche.}

We model an observed burst as a radiative episode with an effective avalanche size \(\mathcal{N}\), defined as the number of participating reconnection units, or equivalently the effective participating length or volume of one burst. The observed spectral energy density is written as
\begin{equation}
	E_\nu=\epsilon_\nu \mathcal{N}^{D_{\rm eff}},
\end{equation}
where \(\epsilon_\nu\) is the characteristic spectral energy density of a localized elementary reconnection episode, and \(D_{\rm eff}\) is an effective mapping index that absorbs geometric scaling, coherent radiative efficiency, beaming, bandwidth and propagation effects. This relation is a phenomenological mapping from avalanche size to observed radio spectral energy density, rather than a detailed emission model. Different bursts need not be causally connected to one another; they are treated as independent realizations drawn from the same underlying distribution of avalanche sizes. Similar avalanche statistics have long been invoked in solar-flare energy distributions \cite{Lu1991}.

\paragraph{Subcritical branching and the exponential component.}

A minimal mathematical description of localized reconnection is a branching process. Suppose that one elementary reconnection unit triggers \(k\) secondary units, with \(k\) drawn from a Poisson distribution
\begin{equation}
	p_k=\frac{\mu^k e^{-\mu}}{k!},
\end{equation}
where \(\mu\equiv\mu_{\rm eff}\) is the effective branching ratio, namely the average number of secondary reconnection units triggered by one active unit. It is set by local magnetic stress, current-sheet geometry, magnetization, pair loading, radiative losses and boundary conditions.

For a Galton--Watson process with Poisson offspring, the probability that the total avalanche size is \(\mathcal{N}=n\) is the Borel distribution \cite{Otter1949,Harris1963},
\begin{equation}
	P(\mathcal{N}=n|\mu)=
	\frac{(\mu n)^{n-1}e^{-\mu n}}{n!},
	\qquad n=1,2,3,\dots .
\end{equation}
Using Stirling's approximation,
\begin{equation}
	n!\simeq \sqrt{2\pi}\,n^{n+1/2}e^{-n},
\end{equation}
the large-\(n\) asymptotic form becomes
\begin{equation}
	P(\mathcal{N}=n|\mu)
	\simeq
	\frac{1}{\sqrt{2\pi}\mu}
	n^{-3/2}
	\exp[-n(\mu-1-\ln\mu)] .
\end{equation}
For \(\mu<1\), the factor
\begin{equation}
	a(\mu)\equiv \mu-1-\ln\mu
\end{equation}
is positive, and the avalanche-size distribution has an exponential cutoff,
\begin{equation}
	P(\mathcal{N}|\mu)\propto \mathcal{N}^{-3/2}
	\exp\left[-a(\mu)\mathcal{N}\right].
\end{equation}
Equivalently, the characteristic cutoff size is
\begin{equation}
	\mathcal{N}_0(\mu)=\frac{1}{\mu-1-\ln\mu}.
\end{equation}
For a fixed subcritical \(\mu<1\), localized reconnection episodes terminate before they grow into macroscopic avalanches and produce an avalanche-size distribution dominated by an exponential cutoff. For an approximately linear low-energy mapping \(E_\nu\propto\mathcal{N}\), the cumulative distribution is then well approximated by
\begin{equation}
	N_{\rm exp}(>E_\nu)\propto
	\exp\left(-\frac{E_\nu}{E_0}\right).
\end{equation}
Thus, the exponential component reflects localized subcritical propagation rather than globally plasmoid-stable current sheets.

\paragraph{A distribution of effective branching ratios.}

The coexistence of the exponential component and the power-law tail suggests that \(\mu_{\rm eff}\) should not be regarded as a single source-wide constant. Different local triggers can sample different effective branching states, set by current-sheet geometry, magnetic stress, guide field, pair loading and boundary conditions. The observed avalanche-size distribution is therefore better viewed as a mixture over effective branching ratios,
\begin{equation}
	P(\mathcal{N})
	=
	\int d\mu\,
	p(\mu)\,
	P(\mathcal{N}|\mu).
	\label{eq:mu_mixture}
\end{equation}
In this picture, the low-energy exponential component is controlled by the subcritical bulk of \(p(\mu)\), while the high-energy power-law tail is controlled by its near-critical tail. This avoids the need to assume that one global \(\mu_{\rm eff}\) jumps between two discrete values. The temporal evolution can be expressed schematically as
\begin{equation}
	p(\mu,t)
	=
	w_{\rm sub}(t) f_{\rm sub}(\mu)
	+
	w_{\rm cr}(t) f_{\rm cr}(\mu),
	\label{eq:mu_decomposition}
\end{equation}
where \(f_{\rm sub}\) is the subcritical bulk responsible for the exponential component, \(f_{\rm cr}\) is the near-critical tail responsible for the power-law component, and \(w_{\rm sub}\), \(w_{\rm cr}\) are their time-dependent occurrence weights.

In this framework, \(E_0\) is associated with the subcritical bulk \(f_{\rm sub}(\mu)\) and the local energy scale of the localized branch. Its observed stability should therefore not be interpreted as evidence that all bursts share a single narrowly distributed \(\mu_{\rm eff}\). Instead, it motivates the quantitative constraints on the subcritical bulk and the local magnetospheric reservoir developed in Supplementary Note 3.

\paragraph{Near-critical tail and the origin of the power law.}

The power-law tail can arise if the event-to-event distribution \(p(\mu)\) extends continuously toward the critical point. Let
\begin{equation}
	\delta=1-\mu .
\end{equation}
Near \(\mu=1\),
\begin{equation}
	a(\mu)=\mu-1-\ln\mu
	\simeq
	\frac{\delta^2}{2}.
\end{equation}
Suppose the near-critical part of the branching-ratio distribution behaves as
\begin{equation}
	p(\mu)d\mu\propto \delta^\beta d\delta .
\end{equation}
At large \(\mathcal{N}\), Equation~(\ref{eq:mu_mixture}) gives
\begin{align}
	P(\mathcal{N})
	&\propto
	\mathcal{N}^{-3/2}
	\int_0^\infty
	\delta^\beta
	\exp\left(-\frac{\mathcal{N}\delta^2}{2}\right)d\delta
	\nonumber\\
	&\propto
	\mathcal{N}^{-\left(2+\beta/2\right)} .
	\label{eq:mu_tail}
\end{align}
Thus, a power-law tail does not require all high-energy events to sit exactly at \(\mu_{\rm eff}=1\). It can emerge if the event-to-event distribution of effective branching ratios has support close to the critical point.

Transforming from avalanche size to observed spectral energy density,
\begin{equation}
	E_\nu=\epsilon_\nu \mathcal{N}^{D_{\rm eff}},
\end{equation}
gives
\begin{equation}
	\mathcal{N}=\left(\frac{E_\nu}{\epsilon_\nu}\right)^{1/D_{\rm eff}},
	\qquad
	\frac{d\mathcal{N}}{dE_\nu}\propto
	E_\nu^{1/D_{\rm eff}-1}.
\end{equation}
Therefore,
\begin{equation}
	p(E_\nu)dE_\nu=P(\mathcal{N})d\mathcal{N}.
\end{equation}
Assuming a general scale-free avalanche-size distribution
\begin{equation}
	P(\mathcal{N})\propto \mathcal{N}^{-\tau_{\mathcal{N}}},
\end{equation}
one obtains
\begin{equation}
	p(E_\nu)\propto
	E_\nu^{-\left[1+\frac{\tau_{\mathcal{N}}-1}{D_{\rm eff}}\right]}.
\end{equation}
The cumulative distribution is
\begin{equation}
	N_{\rm PL}(>E_\nu)\propto E_\nu^{-\gamma},
\end{equation}
with
\begin{equation}
	\gamma=\frac{\tau_{\mathcal{N}}-1}{D_{\rm eff}}.
\end{equation}
For the near-critical mixture above, \(\tau_{\mathcal{N}}=2+\beta/2\). In practice, \(\tau_{\mathcal{N}}\) and \(D_{\rm eff}\) are effective parameters because the observed radio spectral energy density depends on coherent-emission efficiency, beaming, bandwidth, and propagation effects.

\paragraph{Supercritical excursions and finite-size cutoffs.}

If \(\mu>1\), an infinite branching process has a non-zero probability of runaway growth. The extinction probability \(q\) satisfies
\begin{equation}
	q=G(q),
\end{equation}
where \(G(z)\) is the offspring generating function. For Poisson offspring,
\begin{equation}
	G(z)=\exp[\mu(z-1)],
	\qquad
	q=\exp[\mu(q-1)] .
\end{equation}
When \(\mu>1\), \(q<1\), and the runaway probability is \(1-q\). In a finite magnetosphere this branch cannot grow indefinitely. It is limited by the finite current-sheet length, available magnetic free energy, and radiative transparency. Thus, supercritical excursions are expected to appear as rare system-scale events, saturation, or a high-energy cutoff, rather than as an indefinitely extended power law.

\section*{Supplementary Note 3. Magnetospheric constraints from the stable exponential scale \(E_0\)}\label{supsec:E0cons}

The exponential e-folding scale \(E_0\) is used here as the observational anchor for the localized branch of the energy hierarchy. In the framework developed in Supplementary Note 2, \(E_0\) characterizes the subcritical component rather than the near-critical power-law tail. Its stability therefore constrains two related quantities: the local magnetospheric energy reservoir and, in a limiting branching interpretation, the subcritical bulk of the effective branching-ratio distribution.

\paragraph*{Magnetic-energy scaling.}

To connect the observed characteristic spectral energy density with the magnetar's physical state, we start from the local magnetic field given in Equation~(\ref{eq:B_loc}). Since tearing instabilities tap primarily into the non-potential twisted component of the field, we parameterize the effective reconnecting field as
\begin{equation}
	B_{\rm rec}=\kappa_B B_{\rm loc},
\end{equation}
where \(0<\kappa_B\le1\). In what follows, the subscript ``0'' denotes quantities associated with the localized component characterized by \(E_0\). The magnetic free-energy density available for dissipation is therefore
\begin{equation}
	u_{\rm free}\simeq
	\frac{B_{\rm rec}^2}{8\pi}
	=
	\frac{\kappa_B^2B_s^2}{8\pi}
	\left(\frac{R_*}{R_{\rm em}}\right)^6 .
	\label{eq:ufree}
\end{equation}
For a localized effective active volume
\begin{equation}
	V_0=\xi_{0,\rm eff}R_{\rm em}^3,
\end{equation}
where \(\xi_{0,\rm eff}\) is a radio-weighted effective active-volume factor that absorbs both the elementary active volume and the finite subcritical propagation scale, the true emitted radio energy associated with the exponential scale is
\begin{equation}
	E_{\rm true,0}
	\simeq
	\epsilon_{\rm rad,0}u_{\rm free}V_0
	\simeq
	\epsilon_{\rm rad,0}
	\frac{\kappa_{B,0}^2B_s^2R_*^6}{8\pi R_{\rm em}^3}
	\xi_{0,\rm eff}.
	\label{eq:etrue_e0}
\end{equation}
Assuming that the radio emission is relativistically beamed into a solid angle \(\Delta\Omega\sim\pi/\Gamma_0^2\), the isotropic-equivalent energy is
\begin{equation}
	E_{\rm iso,0}
	\simeq
	\frac{4\pi}{\Delta\Omega}E_{\rm true,0}
	\simeq
	4\Gamma_0^2E_{\rm true,0}.
\end{equation}
Dividing by the effective bandwidth \(\Delta\nu_{\rm eff}\) gives the observed isotropic-equivalent spectral energy density,
\begin{equation}
	E_0
	\simeq
	\frac{\Gamma_0^2\epsilon_{\rm rad,0}\kappa_{B,0}^2\xi_{0,\rm eff}}
	{2\pi\Delta\nu_{\rm eff}}
	\frac{B_s^2R_*^6}{R_{\rm em}^3}.
	\label{eq:E0_master}
\end{equation}
This is the master relation used below, with the stable exponential scale \(E_0\) serving as the observational anchor.

\paragraph*{Constraints on \(R_{\rm em}\) and \(B_s\).}

Introducing
\begin{equation}
	B_s=10^{15}B_{15}\,{\rm G},\qquad
	R_*=10^6R_6\,{\rm cm},\qquad
	R_{\rm em}=10^8r_8\,{\rm cm},
\end{equation}
\begin{equation}
	\Delta\nu_{\rm eff}=10^9\Delta\nu_9\,{\rm Hz},\qquad
	\Gamma_0=10\Gamma_{0,1},\qquad
	\epsilon_{\rm rad,0}=10^{-3}\epsilon_{0,-3},
\end{equation}
\begin{equation}
	\kappa_{B,0}^2=0.1\kappa_{0,-1}^2,\qquad
	\xi_{0,\rm eff}=0.1\xi_{0,-1},
\end{equation}
and defining
\begin{equation}
	\Lambda_0\equiv
	\Gamma_{0,1}^2\epsilon_{0,-3}
	\kappa_{0,-1}^2\xi_{0,-1}
	R_6^6\Delta\nu_9^{-1},
	\label{eq:def_lambda0}
\end{equation}
Equation~(\ref{eq:E0_master}) becomes
\begin{equation}
	E_0
	\simeq
	1.59\times10^{29}
	\Lambda_0
	B_{15}^2r_8^{-3}
	\ {\rm erg\,Hz^{-1}}.
	\label{eq:E0_scaled}
\end{equation}
For the full FAST sample, \(E_0=5.68\times10^{28}\,{\rm erg\,Hz^{-1}}\), giving
\begin{equation}
	\Lambda_0B_{15}^2r_8^{-3}
	\simeq
	0.36 .
\end{equation}
Equivalently,
\begin{equation}
	R_{\rm em}
	\simeq
	1.41\times10^8
	\Lambda_0^{1/3}B_{15}^{2/3}
	\left(
	\frac{E_0}{5.68\times10^{28}\ {\rm erg\,Hz^{-1}}}
	\right)^{-1/3}
	{\rm cm},
	\label{eq:E0_Rem}
\end{equation}
and
\begin{equation}
	B_s
	\simeq
	5.97\times10^{14}
	\Lambda_0^{-1/2}r_8^{3/2}
	\left(
	\frac{E_0}{5.68\times10^{28}\ {\rm erg\,Hz^{-1}}}
	\right)^{1/2}
	{\rm G}.
	\label{eq:E0_Bs}
\end{equation}
For canonical magnetar surface fields and \(\Lambda_0\sim1\), the inferred radius is \(R_{\rm em}\sim10^8\,{\rm cm}\), placing the localized dissipation layer in the inner-to-middle magnetosphere. The inferred radius depends only weakly on the uncertain nuisance parameters,
\(R_{\rm em}\propto\Lambda_0^{1/3}\). Therefore, reasonable variations in beaming, radiative efficiency, reconnecting-field fraction, bandwidth, or effective active volume do not qualitatively alter the above conclusion.

Equivalently, Equation~(\ref{eq:E0_master}) gives a joint constraint on the reconnecting-field fraction and the effective active-volume factor,
\begin{equation}
	\kappa_{B,0}^2\xi_{0,\rm eff}
	\simeq
	3.6\times10^{-3}
	\Gamma_{0,1}^{-2}
	\epsilon_{0,-3}^{-1}
	B_{15}^{-2}
	R_6^{-6}
	r_8^3
	\Delta\nu_9
	\left(
	\frac{E_0}{5.68\times10^{28}\ {\rm erg\,Hz^{-1}}}
	\right).
	\label{eq:kappa_xi_e0}
\end{equation}
For fiducial parameters, this implies
\(\kappa_{B,0}^2\xi_{0,\rm eff}\sim3.6\times10^{-3}\). If
\(\kappa_{B,0}^2\sim0.1\), the corresponding effective active-volume factor is
\(\xi_{0,\rm eff}\sim3.6\times10^{-2}\). This value should not be interpreted as the geometric volume fraction of a single microscopic tearing cell; \(\xi_{0,\rm eff}\) is a radio-weighted effective factor that absorbs the active volume and the finite subcritical propagation scale. Even so, it is many orders of magnitude larger than the minimal microscopic expectation
\(\sim S_c^{-2}\sim10^{-8}\) estimated from the aspect-ratio scaling of a single critical plasmoid layer with \(S_c\sim10^4\), supporting the view that the observed low-energy scale is associated with a mesoscopic reconnecting region rather than an isolated microscopic dissipation site.

\paragraph*{Logarithmic stability budget.}

The stability of \(E_0\) constrains the product
\begin{equation}
	E_0\propto
	\Gamma_0^2\epsilon_{\rm rad,0}\kappa_{B,0}^2\xi_{0,\rm eff}
	B_s^2R_{\rm em}^{-3}\Delta\nu_{\rm eff}^{-1}.
\end{equation}
Taking logarithmic variations between two epochs gives
\begin{equation}
	\Delta\ln E_0
	=
	2\Delta\ln\Gamma_0
	+\Delta\ln\epsilon_{\rm rad,0}
	+2\Delta\ln\kappa_{B,0}
	+\Delta\ln\xi_{0,\rm eff}
	+2\Delta\ln B_s
	-3\Delta\ln R_{\rm em}
	-\Delta\ln\Delta\nu_{\rm eff}.
	\label{eq:E0_log_budget}
\end{equation}
For the same observing band, and given that \(B_s\) cannot evolve appreciably over 3.2 years, this reduces to
\begin{equation}
	\Delta\ln E_0
	\simeq
	2\Delta\ln\Gamma_0
	+\Delta\ln\epsilon_{\rm rad,0}
	+2\Delta\ln\kappa_{B,0}
	+\Delta\ln\xi_{0,\rm eff}
	-3\Delta\ln R_{\rm em}.
\end{equation}
Across the four epochs,
\begin{equation}
	\frac{E_{0,\max}}{E_{0,\min}}
	=
	\frac{6.82}{4.83}
	\simeq1.41,
	\qquad
	\Delta\ln E_0\simeq0.34 .
\end{equation}
This gives a simple stability budget. If only one parameter varied at a time, the maximum allowed variation would be approximately
\begin{equation}
	\begin{array}{c|c}
		\hbox{quantity varied alone} & \hbox{maximum allowed factor} \\
		\hline
		\Gamma_0 & 1.41^{1/2}\simeq1.19 \\
		\kappa_{B,0} & 1.41^{1/2}\simeq1.19 \\
		\epsilon_{\rm rad,0} & 1.41 \\
		\xi_{0,\rm eff} & 1.41 \\
		R_{\rm em} & 1.41^{1/3}\simeq1.12
	\end{array}
\end{equation}
Thus, if the microphysical factors remain approximately stable, the characteristic emission radius can drift by at most \(\sim10\%\). Conversely, if the emission radius is fixed, the combined microphysical factor $\Gamma_0^2\epsilon_{\rm rad,0}\kappa_{B,0}^2\xi_{0,\rm eff}$ can vary by no more than a factor of \(\sim1.4\). Large, uncorrelated changes in beaming, reconnecting-field fraction, radiative efficiency or active volume would therefore wash out the observed stability of \(E_0\), unless compensated by correlated changes in other parameters.

\paragraph*{Connection to the branching-ratio distribution.}

The preceding stability budget treats \(\xi_{0,\rm eff}\) as an effective phenomenological factor. If part of this factor reflects the finite propagation scale of the subcritical branch, we may write
\begin{equation}
	\xi_{0,\rm eff}\propto \mathcal{N}_{\rm sub}^{D_0},
\end{equation}
where \(\mathcal{N}_{\rm sub}\) is the characteristic participating size of the subcritical bulk of \(p(\mu)\). In the limiting case where the low-energy branch is represented by a single effective subcritical branching ratio, one has
\begin{equation}
	\mathcal{N}_{\rm sub}\sim
	\mathcal{N}_0(\mu_{\rm eff})
	=
	\frac{1}{\mu_{\rm eff}-1-\ln\mu_{\rm eff}} .
\end{equation}
If the temporal variation of \(E_0\) were dominated by a drift in this effective branching ratio, then
\begin{equation}
	\Delta\ln E_0
	\simeq
	D_0\,\Delta\ln\mathcal{N}_0
	=
	D_0
	\frac{1/\mu_{\rm eff}-1}
	{\mu_{\rm eff}-1-\ln\mu_{\rm eff}}
	\Delta\mu_{\rm eff}.
	\label{eq:E0_mu_variation}
\end{equation}
This expression shows that the low-energy scale becomes increasingly sensitive to small changes in \(\mu_{\rm eff}\) as the subcritical branch approaches the critical point.

Using the same observed range, \(\Delta\ln E_0\approx0.34\), this limiting case gives the following bounds on the allowed drift of the effective subcritical branching state:
\begin{equation}
	\begin{array}{c|c|c}
		\mu_{\rm eff} &
		\displaystyle
		\frac{1/\mu_{\rm eff}-1}
		{\mu_{\rm eff}-1-\ln\mu_{\rm eff}}
		&
		\displaystyle
		|\Delta\mu_{\rm eff}|_{\rm max}
		\\
		\hline
		0.5 & 5.2 & 0.07/D_0 \\
		0.7 & 7.6 & 0.05/D_0 \\
		0.8 & 10.8 & 0.03/D_0 \\
		0.9 & 20.7 & 0.02/D_0
	\end{array}
\end{equation}
If \(E_0\) were controlled mainly by the branching cutoff, the subcritical bulk of the branching-ratio distribution could not drift substantially during the monitoring baseline, especially if it lies close to the critical point. This limiting calculation should not be read as evidence that all bursts share one narrowly varying \(\mu_{\rm eff}\). Rather, the stability of \(E_0\) constrains the low-energy branch of the hierarchy, not a single fixed branching ratio across the full burst population.

\section*{Supplementary Note 4. Constraints from the high-energy power-law tail}

The high-energy tail provides information complementary to the stable exponential scale \(E_0\). Whereas \(E_0\) anchors the characteristic energy scale of localized subcritical reconnection episodes and provides the primary magnetospheric constraint, the observed power-law index \(\gamma\) constrains how the near-critical avalanche branch is projected into the observed radio band. In the notation of Supplementary Note 2, this tail corresponds to the near-critical part of the branching-ratio distribution \(p(\mu)\).

We use the avalanche-size mapping derived in Supplementary Note 2 rather than re-deriving it here. In this framework, each high-energy burst is treated as an independent realization of a reconnection avalanche with an effective participating size \(\mathcal{N}\). The population-wide avalanche-size index \(\tau_{\mathcal{N}}\), the effective projection index \(D_{\rm eff}\), and the observed cumulative power-law slope \(\gamma\) are related by
\begin{equation}
	\gamma=\frac{\tau_{\mathcal{N}}-1}{D_{\rm eff}},
	\qquad
	\tau_{\mathcal{N}}=1+\gamma D_{\rm eff}.
	\label{eq:gamma_tauN_deff}
\end{equation}
Here \(\tau_{\mathcal{N}}\) is the index of the population-wide distribution of effective avalanche sizes, not the instantaneous plasmoid-size distribution within a single reconnecting layer. The index \(D_{\rm eff}\) is likewise an effective projection parameter rather than a purely geometric dimension; it absorbs coherent-emission efficiency, beaming, bandwidth, plasma-transparency and propagation effects.

For the full FAST sample ($\gamma = 0.85$), adopting illustrative geometric limits with $D_{\rm eff}=2$ (area-like scaling) or $D_{\rm eff}=3$ (volume-like scaling) yields $\tau_{\mathcal{N}}\simeq 2.70$ or $3.55$, respectively. Alternatively, if \(\tau_{\mathcal{N}}\sim2\) is adopted as a representative reference value for avalanche-like hierarchical statistics motivated by plasmoid-mediated reconnection \cite{Uzdensky2010,HuangBhattacharjee2012,Takamoto2013,LoureiroUzdensky2016,Sironi2016}, the observed tail implies $D_{\rm eff}\simeq1.18$. This would correspond to a shallower-than-area effective mapping between the cascade scale and the observed radio spectral energy density, as may occur if the escaping coherent radio emission is dominated by localized, highly radiative regions rather than by the full dissipating magnetic volume.

This mapping should be regarded as phenomenological rather than calorimetric. FRBs are coherent radio bursts, and their observed spectral energy density need not scale linearly with the total dissipated magnetic energy. Thus, we do not assume that the coherent radio bursts inherit the same
power-law index as the underlying magnetic-energy cascade. The key requirement is only that the high-energy bursts sample a scale-free range of effective participating reconnection scales, with the detailed conversion into escaping GHz radiation remaining model dependent \cite{Philippov2019,Beloborodov2020,Lyubarsky2020,Mahlmann2022,Wang2023}.

As noted in Supplementary Note 1, the epoch-resolved values of \(\gamma\) are broadly consistent during the burst-rich Epochs 1--3, while the flatter Epoch 4 value is likely affected by limited high-energy statistics. We therefore use the full-sample value as the fiducial characterization of the source-averaged high-energy cascade.

Together with the stable \(E_0\), the high-energy tail is consistent with a picture in which FRB 20220529 repeatedly samples a stable localized dissipation scale while still accessing a near-critical range of avalanche sizes. This interpretation remains phenomenological, but it provides a compact way to connect the invariant low-energy scale and the scale-free bright-end tail within the same reconnection hierarchy.

\section*{Supplementary Note 5. Physical and observational origin of morphological diversity in burst energy distributions}

\subsubsection*{Reconciling phenomenological diversity}

The cumulative burst-energy distributions reported for repeating FRBs exhibit a wide range of empirical forms, including single power laws, double power laws, and power laws with high-energy exponential cutoffs. Although this diversity may reflect genuine source-to-source differences, part of it can also arise naturally within the thresholded dissipation hierarchy described above. In this view, the observed morphology is shaped by both the intrinsic magnetospheric state of the source and the observational window through which the burst population is sampled.

\subsubsection*{Observational selection effects and dynamic range}

Resolving an exponential-plus-power-law (EXP+PL) structure requires a broad energy range, high burst statistics, and a well-characterized completeness threshold. These conditions are most clearly satisfied by FRB~20220529 and, to a lesser extent, FRB~20220912A, whose published sample contains hundreds of bursts spanning several orders of magnitude in spectral energy density. More specifically, three observational requirements must be met. First, the data must reach below or close to the characteristic scale \(E_0\). If the completeness threshold lies above \(E_0\), the low-energy exponential component will be hidden below the detection limit, leaving only the scale-free bright-end tail visible; such a dataset may appear consistent with a single power law even if a characteristic low-energy component is present intrinsically. Second, the sample must contain enough bursts over a sufficiently broad energy range to resolve the smooth curvature between the localized and avalanche-like branches. If the sample contains too few bright bursts, or covers only a narrow range around the transition, a broken or double power law may provide an adequate empirical approximation without implying two independent scale-free states. Third, competing empirical descriptions must be compared under the same completeness threshold and fitting procedure. For FRB~20220529, where these requirements are best satisfied, the EXP+PL form is statistically preferred over the tested alternatives.

High-energy cutoffs may also have multiple origins. They may reflect the finite energy budget of a particular activity episode, limited sampling of rare bright bursts, or a physical upper bound on the maximum participating reconnection scale. In FRB~20220529, by contrast, the bright-end tail persists over a multi-year baseline and is independently supported by the Parkes verification sample. The observed EXP+PL morphology is therefore better interpreted as a stable energy-release hierarchy than as the transient cutoff of a single activity episode.

\subsubsection*{Intrinsic physical drivers of component visibility}

The characteristic exponential scale \(E_0\) is not expected to be universal across repeating FRBs. As discussed in Supplementary Note 3, \(E_0\) depends on the local magnetic energy reservoir and on effective radiative and geometric factors. Variations in surface field strength, emission radius, radiative efficiency, beaming, or participating volume can therefore shift the localized component to different observed energy ranges.

The relative visibility of the exponential and power-law branches may also depend on activity state. A long-lived, gradually fading source can sample both localized reconnection episodes and near-critical avalanches over an extended baseline, allowing the full EXP+PL morphology to emerge. During a short, highly active outburst, by contrast, the detected sample may be dominated by larger participating regions, making the distribution appear more power-law-like. This provides one possible interpretation of why some active repeaters, such as FRB 20201124A during enhanced-activity phases, do not presently show a resolved low-energy exponential component. This interpretation remains tentative and requires longer, more uniformly selected monitoring data.

\subsubsection*{Implications for population-level comparisons}

The similarity between FRB 20220529 and FRB 20220912A suggests that a characteristic low-energy component coexisting with a scale-free bright-end tail can recur in long-lived hyperactive repeaters. However, the current comparison does not establish universality across the repeating-FRB population. For several other active repeaters, existing samples remain inconclusive because they lack sufficient burst numbers, dynamic range, cadence uniformity, or completeness characterization to resolve both sides of the putative transition.

We therefore interpret the apparent diversity of published energy distributions as arising from a combination of observational selection and intrinsic magnetospheric state. The absence of a resolved transition in some sources should not be regarded as a population-level counterexample unless future datasets with adequate sensitivity, dynamic range, and temporal coverage still fail to reveal a stable two-component hierarchy. Long-term, high-cadence, and wide-band monitoring will be essential for determining whether stable two-component energy hierarchies are common among active repeating FRBs or restricted to a subset of long-lived hyperactive engines.

%\setcounter{figure}{0} % Reset figure counter
%\captionsetup[figure]{name={\bf Extended Data Figure}}

\begin{figure*}
	\centering
	\includegraphics[width=\textwidth]{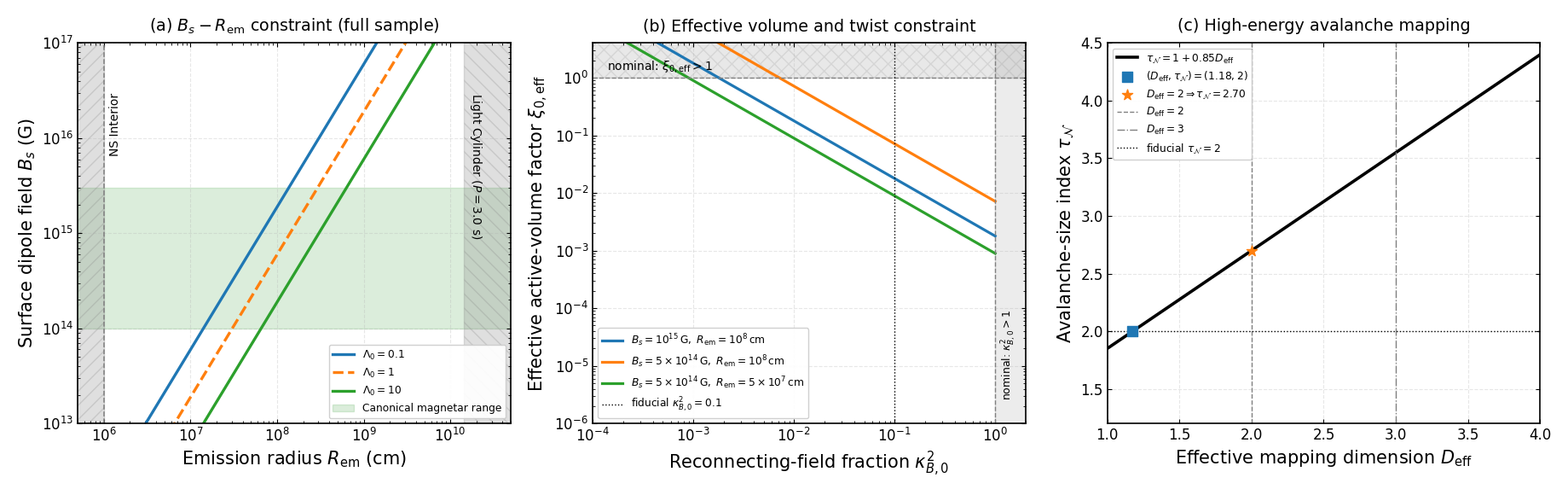}
	\caption{\textbf{Full-sample magnetar-parameter constraints derived from the stable exponential scale \(E_0\).}
		\textbf{(a)} Locus in the \(B_s\)--\(R_{\rm em}\) plane implied by the full-sample \(E_0\). The three curves correspond to different values of the composite parameter \(\Lambda_0\) (Equation~\ref{eq:def_lambda0}), which encapsulates beaming, radiative efficiency, reconnecting-field fraction, effective active volume, bandwidth, and neutron-star radius. The shaded bands indicate canonical magnetar surface fields.
		\textbf{(b)} Constraints on the effective active-volume factor \(\xi_{0,\rm eff}\) as a function of the reconnecting-field fraction \(\kappa_{B,0}^2\). Shaded regions denote the nominally disfavored regime for a pure volume-fraction interpretation (\(\xi_{0,\rm eff}>1\) or \(\kappa_{B,0}^2>1\)).
		\textbf{(c)} Phenomenological mapping \(\tau_{\mathcal{N}}=1+\gamma D_{\rm eff}\) for the high-energy tail, relating the observed cumulative power-law index \(\gamma\) to the population-wide avalanche-size index \(\tau_{\mathcal{N}}\) and the effective projection index \(D_{\rm eff}\).}
	\label{fig:sup_fullsample_constraints}
\end{figure*}

\begin{figure*}
	\centering
	\includegraphics[width=\textwidth]{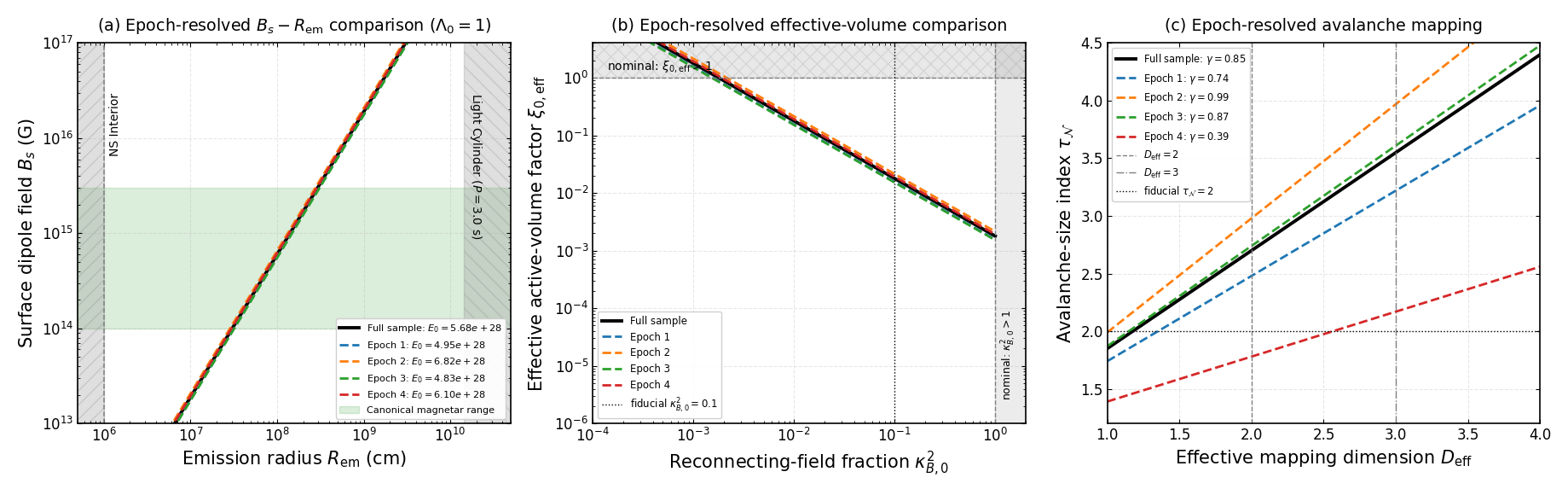}
	\caption{\textbf{Epoch-resolved robustness test of the \(E_0\)-based magnetar-parameter constraints.}
		\textbf{(a)} Comparison of the inferred \(B_s\)--\(R_{\rm em}\) loci across the four epochs and the full sample, calculated for a fixed \(\Lambda_0=1\) to highlight the stability of the \(E_0\)-anchored constraint.
		\textbf{(b)} Constraints on \(\xi_{0,\rm eff}\) versus \(\kappa_{B,0}^2\) for a fiducial magnetar configuration (\(B_{15}=1\), \(r_8=1\)).
		\textbf{(c)} Epoch-resolved mapping between the high-energy slope and the avalanche hierarchy. Epochs 1--3 remain broadly consistent with the full-sample inference, while the shift in Epoch 4 reflects the statistically flattened late-time tail in the low-activity regime.}
	\label{fig:sup_epoch_constraints}
\end{figure*}

%% METHODS REFERENCES
\clearpage
\noindent {\bf References} \\

\clearpage

\subsubsection*{Data availability}

\noindent
Raw data are available from the FAST data center, \url{http://fast.bao.ac.cn}. Calibrated data are available upon request.

\subsubsection*{Code availability}
\noindent
\textsc{PRESTO} (\url{http://www.cv.nrao.edu/~sransom/presto/})

\noindent
\textsc{HEIMDALL} (\url{https://sourceforge.net/projects/heimdall-astro/})

\noindent
\textsc{PSRCHIVE} (\url{http://psrchive.sourceforge.net})

\noindent Custom scripts used for burst-energy fitting and figure generation are available from the corresponding authors upon reasonable request.

\begin{addendum}

\item This work is partially supported by the National Natural Science Foundation of China (grant Nos. 12321003, 12393813, 12373052, 12273113, 12233002, 12003028, 12503058), CAS Project for Young Scientists in Basic Research (YSBR-063), the National Key R\&D Program of China (2021YFA0718500), Postdoctoral Fellowship Program of CPSF (grant No. GZC20252100), the ACAMAR Postdoctoral Fellow, China Postdoctoral Science Foundation (grant No. 2025M773201), and Jiangsu Funding Program for Excellent Postdoctoral Talent. \\

\item[Author Contributions]  X.Y. and S.B.Z. led the radio data analyses. D.X. and X.F.W. led the interpretation. X.Y., D.X., S.B.Z., W.L.Z., and X.F.W. led the manuscript writing. Y.L., S.B.Z. and X.F.W. coordinated the observing campaign and cosupervised the data analysis. 
Y.P.Y., J.J.W., J.S.W, J.J.G., F.Y.W. and Z.G.D. cosupervised the data interpretation.
W.L.Z., X.F.W., and X.Y. contributed to the theoretical and statistical analysis for the data fitting.

\item[Competing Interests] The authors declare that they have no competing financial interests.

\item[Correspondence] Correspondence and requests for materials should be addressed to D. Xiao, X. F. Wu or Z. G. Dai.

\end{addendum}
\clearpage

\end{document}